\documentclass[10pt]{article}
\usepackage{amsmath}     
\usepackage{amssymb}    
\usepackage{epsfig}  
\usepackage{sparticles}
\usepackage{feynmf}

\def \BE {\begin{equation}}
\def \EE {\end{equation}}
\def \BEA {\begin{eqnarray}}
\def \EEA {\end{eqnarray}}
\def \CR {\nonumber \\}
\def \zer {^{(0)}}
\def \one {^{(1)}}
\def \two {^{(2)}}

\def \e {{\epsilon}}

\begin{document}
\title{ Wave Turbulence}
\author{Yeontaek Choi$^*$, Yuri V. Lvov$^\dagger$, Sergey Nazarenko$^*$\\
\\
$^*$ Mathematics Institute,
The University of Warwick, 
\\
Coventry, CV4-7AL, UK
\\
$^\dagger$ Department of Mathematical Sciences, Rensselaer
Polytechnic Institute, 
\\
Troy, NY 12180 
}

\maketitle

\abstract{In this paper we review recent developments in the 
statistical theory of weakly nonlinear dispersive waves, the subject 
known as Wave Turbulence (WT). We revise WT theory using a generalisation
of the random phase approximation (RPA).  This generalisation takes into
account that not only the phases but also the amplitudes of the wave
Fourier modes are random quantities and it is called the
``Random Phase and Amplitude''  approach. This approach allows to
systematically derive the kinetic equation for the energy spectrum from
the the Peierls-Brout-Prigogine (PBP) equation for the multi-mode
probability density function (PDF).  The PBP equation was originally derived for the
three-wave systems and in the present paper we derive a similar equation for
the four-wave case. 
Equation for the multi-mode PDF will be used to validate the
statistical assumptions about the phase and the amplitude randomness used for
WT closures. Further, the multi-mode PDF
contains a detailed statistical information, beyond spectra, and it finally
allows to study non-Gaussianity and intermittency in WT, as it will be
described in the present paper. In particular, we will show that intermittency
of stochastic nonlinear waves is related to a flux of probability in the space
of wave amplitudes. }

\section{Introduction}
Imagine surface waves on sea produced by wind of moderate strength,
so that the surface is smooth and there is no whitecaps.
Typically, these waves exhibit great deal of randomness and
the theory which aims to describe their statistical properties is called
Wave Turbulence (WT). More broadly, WT deals the fields of dispersive
waves which are engaged in stochastic weakly nonlinear interactions
over a wide range of scales in various physical media. 
Plentiful examples of WT are found in
oceans, atmospheres, plasmas and Bose-Einstein
condensates~\cite{ZLF,Ben,GS,Newell68,Zakfil,hasselman,DNPZ,llnz,davidson,ZakharovPRL,jansen}.
WT theory has a long and exciting history which started in 1929
 from the pioneering paper of Peierls who derived a
 kinetic equation for phonons in solids
\cite{peierls}. In the 1960's these ideas have been vigorously developed 
in oceanography \cite{hasselman,Zakfil,Ben,Newell68,jansen} and in 
plasma physics \cite{GS,davidson,lvovzakh}. First of all, both the ocean
and the plasmas can support great many types of dispersive propagating waves,
and these waves play key role in turbulent transport phenomena, particularly
the wind-wave friction in oceans and the anomalous diffusion and
thermo-conductivity in tokamaks.  Thus, WT kinetic equations where developed
and analysed for different types of such waves. A great development in the
general WT theory was done by in the papers of Zakharov and Filonenko
\cite{Zakfil}. Before this work it was generally understood that the nonlinear
dispersive wavefields are statistical, but it was also thought that such a
``gas'' of stochastic waves is close to thermodynamic equilibrium. Zakharov
and Filonenko \cite{Zakfil} were the first to argue that the stochastic
wavefields are more like Kolmogorov turbulence which is determined by the rate
at which energy cascades through scales rather than by a thermodynamic
``temperature'' describing the energy equipartition in the scale space. This
picture was substantiated by a remarkable discovery of an exact solution to
the wave-kinetic equation which describes such Kolmogorov energy
cascade. These solutions are now commonly known as Kolmogorov-Zakharov (KZ) spectra
and they form the nucleus of the WT theory.

Discovery of the KZ spectra was so powerful that it dominated the WT theory
for decades thereafter. Such spectra were found for a large variety of
physical situations, from quantum to astrophysical applications, and a great
effort was put in their numerical and experimental verification.  For a long
time, studies of spectra dominated WT literature. A detailed account of these
works was given in \cite{ZLF} which is the only book so far written on this
subject. Work on these lines has continued till now and KZ spectra were found
in new applications, particularly in astrophysics \cite{gnnp}, ocean interior
\cite{LT} and even cosmology \cite{Micha_Tkachev}. 
However, the spectra do not tell the whole story about the turbulence 
statistics. In particular, they do not tell us if the wavefield statistics is
Gaussian or not and, if not, in what way. This question is of general
importance in the field of Turbulence because it is related with the
intermittency
phenomenon, - an anomalously high probability of large
fluctuations. Such ``bursts'' of turbulent wavefields were predicted
 based on a scaling analysis in \cite{biven}
and they were linked to formation of coherent structures, such as whitecaps on
sea \cite{rough} or collapses in optical turbulence \cite{DNPZ}.
To study these problems qualitatively, the kinetic equation description is not
sufficient and one has to deal directly with the probability density functions (PDF).

In fact, such a description in terms of the PDF appeared already in the the Peierls 1929 paper
simultaneously
with the kinetic equation for waves \cite{peierls}. This result was largely
forgotten by the WT community because fine statistical details and
intermittency
had not interested turbulence researchers until relatively recently and also
because, perhaps, this result got in the shade of the KZ spectrum
discovery. However, this line of investigation was continued by Brout and
Prigogine \cite{bp} who derived an evolution equation for the multi-mode
PDF commonly known as the Brout-Prigogine equation. This approach 
was applied to the study of randomness underlying the
WT closures by Zaslavski and Sagdeev \cite{zs}. All of these authors, Peierls,
Brout and Prigogine and Zaslavski and Sagdeev restricted their consideration
to the nonlinear interaction arising from the potential energy only
(i.e. the interaction Hamiltonian involves coordinates but not
momenta). This restriction leaves out the capillary water waves,
Alfven, internal and Rossby waves, as well as many other interesting
WT systems. Recently, this restriction was removed  by considering the
most general three-wave Hamiltonian systems \cite{physd}. It was
shown that the multi-mode PDF still obeys the Peierls-Brout-Prigogine (PBP)
 equation in this general case. This work will be described in the present review.
We will also present, for the first time, a derivation of the evolution
equation for the multi-mode PDF for the general case of four
wave-systems. This equation is applicable, for example, to WT of the deep
water surface gravity waves and waves in Bose-Einstein condensates or optical media
described by the nonlinear Schroedinger (NLS) equation. 

We will also describe the analysis of papers  \cite{physd} of the
randomness assumptions underlying the statistical WT closures. Previous
analyses in this field examined validity of the random phase assumption 
\cite{bp,zs} without devoting much attention to the amplitude statistics.
Such ``asymmetry'' arised from a common mis-conception that the phases evolve much faster than
amplitudes in the system of nonlinear dispersive waves and, therefore,
the averaging may be made over the phases only ``forgetting'' that the
amplitudes are statistical quantities too (see
e.g. \cite{ZLF}).  This statement become less obvious if one takes
into account that we are talking not about the linear phases $\omega
t$ but about the phases of the Fourier modes in the interaction
representation.  Thus, it has to be the nonlinear frequency
correction that helps randomising the phases \cite{zs}.
On the other hand, for three-wave systems 
the period associated with the nonlinear frequency correction
is of the same $\epsilon^2$ order in small nonlinearity $\epsilon$
as the nonlinear  evolution time and, therefore, phase randomisation
cannot occur faster that the nonlinear evolution of the amplitudes.
One could hope that the situation is better for 4-wave systems
 because the nonlinear frequency correction
is still $\sim \epsilon^2$ but the nonlinear evolution appears
only in the  $\epsilon^4$ order. However, in order to make the
asymptotic analysis consistent, such  $\epsilon^2$ correction
has to be removed from the interaction-representation amplitudes
and the remaining phase and amplitude evolutions are, again,
at the same time scale (now $1/\epsilon^4$).
This picture is confirmed by the numerical simulations of 
the 4-wave systems \cite{zakhpush,cln} which indicate that the
nonlinear phase evolves at the same timescale as the amplitude.
Thus, to proceed theoretically one has to start with phases which
are already random (or almost random) and hope that this randomness
is preserved over the nonlinear evolution time. In most of the 
previous literature prior to  \cite{physd} such preservation was assumed but not proven.
Below, we will describe the analysis of the extent to which
such an assumption is valid made in  \cite{physd}. 

We will also describe the results of 
\cite{ln} who derived the time evolution equation for higher-order 
moments of the Fourier amplitude, and its  application to description of
statistical wavefields with long correlations and associated ``noisiness''
of the energy spectra characteristic to typical laboratory and numerical
experiments.
We will also describe the results of 
\cite{cln} about the  time evolution of the one-mode PDF and their
consequences for the intermittency of stochastic nonlinear wavefields. In
particular, we will discuss the relation between intermittency and the
probability fluxes in the amplitude space.

\section{\label{SettingTheStage} Setting the stage I: Dynamical Equations of motion}
Wave turbulence formulation  deals with a many-wave system with dispersion and
weak nonlinearity. For systematic
derivations one needs to start from Hamiltonian equation of motion.
Here we consider a system of weakly interacting waves  in
a periodic box \cite{ZLF},
\BEA
i\dot c_l &=&\frac{\partial {\cal H}}{\partial \bar c_l}, 
\label{HamiltonianEquationOfMotion}
\EEA
where $c_l$ is often called the field variable. It represents the
amplitude of the interacting plane wave. The Hamiltonian is represented
as an expansion in powers of small amplitude,
\BE {\cal H} = {\cal H}_2 + {\cal H}_3+ {\cal H}_4 +  {\cal H}_5 + \dots, 
\label{HamiltonianExpansion}
\EE
where $H_j$ is a term proportional to product of $j$ amplitudes $c_l$,
\BE
{\cal H}_j  = \sum\limits_{q_1,q_2,q_3, \dots q_n, p_1,p_2,\dots p_m = 1}^{\infty}
\left(
T^{q_1 q_2 \dots q_n}_{p1,p2 \dots p_m} \bar c_{q_1} \bar c_{q_2}\dots \bar c_{q_m} 
c_{p_1} c_{p_2}\dots c_{p_m}+{\rm c.c}.\right), \ \ \ n+m=j
\nonumber\\
\EE
where $q_1,q_2,q_3, \dots q_n$ and $ p_1,p_2,\dots p_m$ are wavevectors on a
$d$-dimensional Fourier space lattice.
Such general $j$-wave Hamiltonian describe the wave-wave interactions
where $n$ waves collide to create $m$ waves. Here $T^{q_1 q_2 \dots
q_n}_{p1,p2 \dots p_m}$ represents the amplitude of the  $n\to m$ process.  In
this paper we are going to consider expansions of Hamiltonians up to
forth order in wave amplitude.

Under rather general conditions the quadratic part of a Hamiltonian,
which correspond to a linear equation of motion, can be diagonalised
to the form
\BE {\cal H}_2 = \sum_{n=1}^\infty \omega_n|c_n|^2.
\label{QuadraticHamiltonian}
\EE
This form of Hamiltonian correspond to  noninteracting (linear) waves. 
First correction to the quadratic Hamiltonian is a cubic
Hamiltonian, which describes the processes of decaying of single wave
into two waves or confluence of two waves into a single one. Such a Hamiltonian has the 
form
\BEA  {\cal H}_3= 
\epsilon
\sum_{l,m,n=1}^\infty  V^l_{mn} \bar c_{l} c_m c_n\delta^l_{m+n}+c.c.,
\nonumber \\
\label{QubicHamiltonian}
\EEA
where $\epsilon \ll 1$ is a formal parameter corresponding to small
nonlinearity
( $\epsilon$ is proportional to the small amplitude whereas $c_n$ is
normalised so that $c_n \sim 1$.)
Most general form of three-wave Hamiltonian would also have terms describing the 
confluence of three waves or spontaneous appearance of three waves out of  vacuum. Such a 
terms would have a form 
$$\sum_{l,m,n=1}^\infty \ U^{lmn} c_{l} c_m c_n\delta_{l+m+n}+c.c.$$
It can be shown however that for systems that are dominated by three-wave 
resonances such terms do not contribute to long term dynamics of systems. 
We therefore choose to omit those terms. 

The most general four-wave Hamiltonian will have $1\to 3$, $3\to 1$, $2\to
2$, $4\to 0 $ and $0\to 4$ terms. Nevertheless $1\to 3$, $3\to 1$,
$4\to 0 $ and $0\to 4$ terms can be excluded from Hamiltonian by
appropriate canonical transformations, so that we limit our consideration to 
only $2\to 2$ terms of  ${\cal H}_4$, namely
\BEA  {\cal H}_4= 
\epsilon^2
\sum_{m,n,\mu,\nu=1}^\infty 
W^{lm}_{\mu\nu} \bar c_{l} \bar c_m c_\mu c_\nu.
\nonumber \\
\label{QuarticHamiltonian}
\EEA
It turns out that generically most of the weakly nonlinear systems can be
separated into two major classes: the ones dominated by three-wave
interactions, so that ${\cal H}_3$ describes all the relevant dynamics
and ${\cal H}_4$ can be neglected, and the systems where the three-wave 
resonance conditions cannot be satisfied, so that the ${\cal
H}_3$ can be eliminated from a Hamiltonian by an appropriate
near-identical canonical transformation \cite{zakhshulman}.
 Consequently, for the purpose
of this paper we are going to neglect either ${\cal H}_3$ or ${\cal
H}_4$, and study the case of resonant three-wave or four-wave
interactions.

Examples of three-wave system include the water surface capillary waves,
internal waves in the ocean and Rossby waves. The most common examples of the four-wave
systems are the surface gravity waves and waves in the NLS model of nonlinear
optical systems and Bose-Einstein condensates.
For reference we will give expressions for the frequencies and the interaction
coefficients corresponding to these examples.

For the {\em capillary waves} we have \cite{ZLF,Zakfil},
\BE
\omega_j = \sqrt{\sigma k^3},
\EE
and
\BE
V^l_{mn} = {1 \over 8 \pi \sqrt{2 \sigma}} (\omega_{l}  \omega_m
\omega_n)^{1/2} 
\times \left[ {L_{k_m, k_n} \over ( k_m k_n)^{1/2} k_l }
-
 {L_{k_l, -k_m} \over ( k_l k_m)^{1/2} k_n }
-
 {L_{k_l, -k_n} \over ( k_l k_n)^{1/2} k_m }
\right],
\EE
where
\BE
 L_{k_m, k_n}  = ({\bf k}_m  \cdot {\bf k}_n) +  k_m k_n
\EE
and $\sigma$ is the surface tension coefficient.

For the {\em Rossby waves} \cite{zakpit,bnaz},
\BE
\omega_j = {\beta k_{jx} \over 1 + \rho^2 k_j^2},
\EE
and
\BE
V^l_{mn} = -{i \beta \over 4 \pi} |  k_{lx}  k_{mx}  k_{nx} |^{1/2} 
\times \left(  { k_{ly} \over 1 + \rho^2 k_l^2} -  { k_{my} \over 1 + \rho^2
  k_m^2}  - { k_{ny} \over 1 + \rho^2 k_n^2} \right),
\EE
where $\beta$ is the gradient of the Coriolis parameter and $\rho$ is the
Rossby deformation radius.

The simplest  expressions  correspond to the {\em NLS waves} \cite{ZMR85,DNPZ},
\BE
\omega_j=|k_j|^2, \hspace{1cm} W^{lm}_{\mu \nu} = 1.
\EE
The  {\em surface gravity waves} are on the other extreme. The 
frequency is $\omega = \sqrt{gk}$ but the matrix element is given by 
notoriously long expressions which can be found in \cite{ZLF,krasitski}.

\subsection{Three-wave case}

When ${\cal H} = {\cal H}_2 + {\cal H}_3$ we have Hamiltonian in a form 
\BEA
{\cal H} &=& \sum_{n=1}^\infty \omega_n|c_n|^2 +\epsilon
\sum_{l,m,n=1}^\infty \left( 
V^l_{mn} \bar c_{l} c_m  c_n\delta^l_{m+n}+c.c.\right).\nonumber
\EEA
Equation of motion $i\dot c_l =\frac{\partial {\cal
H}}{\partial \bar c_l}$ is mostly conveniently represented in the
interaction representation,
\BEA
i \, \dot a_l = \epsilon \sum_{m,n=1}^\infty \left( V^l_{mn} a_{m}
a_{n}e^{i\omega_{mn}^l t} \, \delta^l_{m+n} 
+ 2 \bar{V}^{m}_{ln} \bar a_{n}
a_{m} e^{-i\omega^m_{ln}t } \, \delta^m_{l+n}\right),
\label{Interaction} \EEA 
where $a_j =c_j e^{i \omega_j t}$ is the complex wave amplitude in the interaction
representation,  $l,m,n \in {\cal Z}^d$ are the indices numbering the
wavevectors,
e.g.  $k_m = 2 \pi m/L $, $L $ is the box side length,
$\omega^l_{mn}\equiv\omega_{k_l}-\omega_{k_m}-\omega_{k_m}$ and
$\omega_l=\omega_{k_l}$ is the wave linear dispersion relation.  Here,
$V^l_{mn} \sim 1$ is an interaction coefficient and $\epsilon$ is
introduced as a formal small nonlinearity parameter.

\subsection{Four-wave  case}

Consider a weakly nonlinear wavefield  dominated by the 4-wave interactions, e.g. 
the water-surface gravity waves~\cite{ZLF,hasselman,OsbornePRL,rough},
Langmuir waves in plasmas~\cite{ZLF,GS} and the waves described by the
nonlinear Schroedinger equation~\cite{DNPZ}. 
The a Hamiltonian is given by (in the appropriately chosen variables)
as 
\BEA
{\cal H}={\cal H}_2 + {\cal H}_4
= \sum_{n=1}^\infty \omega_n|c_n|^2 + \epsilon^2
\sum_{m,n,\mu,\nu=1}^\infty 
W^{lm}_{\mu\nu} \bar c_{l} \bar c_m c_\mu c_\nu.
\CR
\label{QuarticHamiltonian2}
\EEA
As in three-wave case the most convenient form of equation of motion
is obtained in interaction representation, $c_l=b_l e^{-i \omega_l t}$,
so that
\begin{equation}
i \dot b_l = \epsilon \sum_{\alpha\mu\nu}
{ W^{l\alpha}_{\mu\nu}}\bar b_\alpha b_\mu b_\nu 
e^{i\omega^{l\alpha}_{\mu\nu}t}
\delta^{l\alpha}_{\mu\nu}
\label{FourWaveEquationOfMotionB}
\end{equation}
where $W^{l \alpha} _{\mu \nu} \sim 1$
 is an interaction coefficient,
$\omega^{l\alpha}_{\mu\nu} =
\omega_l+\omega_{\alpha}-\omega_{\mu}-\omega_{\nu}$.
%
%
  We are going expand
in $\epsilon$ and consider the long-time behaviour of a wave field, but
it will turn out that to do the perturbative expansion in a self-consistent 
manner  we have to renormalise
the frequency of (\ref{FourWaveEquationOfMotionB}) as
\begin{equation}
i \dot a_l = \epsilon \sum_{\alpha\mu\nu}
{ W^{l\alpha}_{\mu\nu}}\bar a_\alpha a_\mu a_\nu 
e^{i\tilde\omega^{l\alpha}_{\mu\nu}t}
\delta^{l\alpha}_{\mu\nu} - \Omega_l a_l,
\label{FourWaveEquationOfMotionA}
\end{equation}
where $a_l = b_l e^{i \Omega_l t} \;\;$, $\tilde \omega^{l\alpha}_{\mu\nu}
 = \omega^{l\alpha}_{\mu\nu}
 +\Omega_l+\Omega_\alpha-\Omega_\mu-\Omega_\nu$, and
\BE
\Omega_l = 2 \epsilon \sum_\mu
 W^{l \mu}_{l \mu}  A_\mu^2 
\label{FrequencyRenormalization}
\EE
 is the nonlinear frequency shift
 arising from self-interactions.
%

\section{Setting the stage II: Statistical setup}

In this section we are going to introduce statistical objects that
shall be used for the description of the wave systems, PDF's and a generating
functional.

\subsection{Probability Distribution Function. }

Let us consider a wavefield $a({\bf x}, t)$ in a periodic cube of with
side $L$ and let the Fourier transform of this field be $a_l(t)$ where
index $l {\in } {\cal Z}^d$ marks the mode with wavenumber $k_l = 2
\pi l /L$ on the grid in the $d$-dimensional Fourier space.  For
simplicity let us assume that there is a maximum wavenumber $k_{max}$
(fixed e.g. by dissipation) so that no modes with wavenumbers greater
than this maximum value can be excited.  In this case, the total
number of modes is $N = (k_{max} / \pi L)^d$. Correspondingly, index
$l$ will only take values in a finite box, $l \in {\cal B}_N \subset
{\cal Z}^d$ which is centred at 0 and all sides of which are equal to
$k_{max} / \pi L = N^{1/3}$.  To consider homogeneous turbulence, the
large box limit $N \to \infty $ will have to be taken.
\footnote{ It is easily to extend the analysis to the infinite Fourier
space, $k_{max} = \infty$.  In this case, the full joint PDF would
still have to be defined as a $N \to \infty$ limit of an $N$-mode PDF,
but this limit would have to be taken in such a way that both
$k_{max}$ and the density of the Fourier modes tend to infinity
simultaneously.}

Let us write the complex $a_l$ as $a_l =A_l \psi_l $ where $A_l$ is a
real positive amplitude and $\psi_l $ is a phase factor which takes
values on ${\cal S}^{1} $, a unit circle centred at zero in the
complex plane. Let us define the $N$-mode joint PDF ${\cal P}^{(N)}$
as the probability for the wave intensities $A_l^2 $ to be in the
range $(s_l, s_l +d s_l)$ and for the phase factors $\psi_l$ to be on
the unit-circle segment between $\xi_l$ and $\xi_l + d\xi_l$ for all
$l \in {\cal B}_N$.  In terms of this PDF, taking the averages will
involve integration over all the real positive $s_l$'s and along all
the complex unit circles of all $\xi_l$'s,
\BEA \langle f\{A^2, \psi \} \rangle
= \left(
 \prod_{ l {\cal 2 B}_N }  \int_{{\cal R}^{+} } ds_l \oint_{{\cal S}^{1} }
|d \xi_l| \right) \;  {\cal P}^{(N)} \{s, \xi \}
f\{s, \xi \} \label{pdfn} \EEA
where notation $f\{A^2,\psi\}$ means that $f$ depends on all $A_l^2$'s
and all $\psi_l $'s in the set $\{A_l^2, \psi_l; l {\cal 2 B}_N \}$
(similarly, $\{s, \xi \}$ means $\{s_l, \psi_l; l \in {\cal B}_N \}$,
etc). The full PDF that contains the complete statistical information
about the wavefield $a({\bf x}, t)$ in the infinite $x$-space can be
understood as a large-box limit
$${\cal P} \{ s_k, \xi_k \}  =  \lim_{N \to \infty}
{\cal P}^{(N)} \{s, \xi \},
$$
i.e. it is a functional acting on the continuous functions
of the wavenumber, $s_k$ and $\xi_k$.
In the the large box
limit there is  a path-integral version of (\ref{pdfn}),
\BE \langle f\{A^2, \psi \} \rangle =
 \int {\cal D}s \oint
|{\cal D} \xi| \;  {\cal P} \{s, \xi \}  f\{s, \xi \}
\label{mean-path} \EE
The full PDF defined above involves all $N$ modes (for either finite
$N$ or in the $N \to \infty$ limit). By integrating out all the
arguments except for chosen few, one can have reduced statistical
distributions. For example, by integrating over all the angles and
over all but $M$ amplitudes,we have an ``$M$-mode'' amplitude PDF,
\BE
{\cal P}^{(M)}_{j_1, j_2, \dots , j_M} = \left(
\prod_{ l \ne j_1, j_2, \dots , j_M }  \int_{{\cal R}^{+} } ds_l 
\prod_{ m {\cal 2 B}_N } 
\oint_{{\cal S}^{1} }
|d \xi_m| \; \right) {\cal P}^{(N)} \{s, \xi \},
\EE
which depends only on the $M$ amplitudes marked by labels
$j_1, j_2, \dots , j_M  {\cal 2 B}_N$.

\subsection{Definition of an ideal RPA field}

Following the approach of \cite{ln,cln}, we now define a ``Random
Phase and Amplitude'' (RPA) field.\footnote{ We keep the same acronym as in
related ``Random Phase Approximation'' but now interpret it differently
because  (i) we emphasise the amplitude randomness and (ii) now RPA is a
defined property of the field to be examined and not an approximation.}
We say that the field $a$ is of
RPA type if it possesses the following statistical properties:

\begin{enumerate}
\item All amplitudes $A_l$ and their phase factors $\psi_l $ are
independent random variables, i.e. their joint PDF is equal to the
product of the one-mode PDF's corresponding to each individual
amplitude and phase,
$$
{\cal P}^{(N)} \{s, \xi \}  = \prod_{ l {\cal 2 B}_N} P^{(a)}_l (s_{l})
P^{(\psi)}_l (\xi_{l})
$$
\item The phase factors $\psi_l$ are uniformly distributed on the unit
circle in the complex plane, i.e.  for any mode $l$
$$
P^{(\psi)}_l (\xi_{l}) = 1/2\pi.
$$
\end{enumerate}
Note that RPA does not fix any shape of the amplitude PDF's and,
therefore, can deal with strongly non-Gaussian wavefields.  Such study
of non-Gaussianity and intermittency of WT was presented in
\cite{ln,cln} and will not be repeated here.  However, we will study
some new objects describing statistics of the phase.

In \cite{ln,cln} RPA was {\em assumed} to hold over the nonlinear
time.  In \cite{physd} this assumption was examined {\em a posteriori},
i.e. based on the evolution equation for the multi-point PDF obtained with RPA
initial fields. Below we will describe this work.
 We will see that RPA fails to hold in its pure form as
formulated above but it survives in the leading order so that the WT
closure built using the RPA is valid.  We will also see that
independence of the the phase factors is quite straightforward,
whereas the amplitude independence is subtle. Namely, $M$ amplitudes
are independent only up to a $O(M/N)$ correction. Based on this
knowledge, and leaving justification for later on in this paper, we
thus reformulate RPA in a weaker form which holds over the nonlinear
time and which involves $M$-mode PDF's with $M \ll N$ rather than the
full $N$-mode PDF.

\subsection{Definition of an essentially RPA field}

We will say that the field $a$ is of an ``essentially RPA'' type if:

\begin{enumerate}

\item The phase factors are statistically independent and uniformly
distributed variables up to  $O(\e^2)$ corrections, i.e.
\BE
{\cal P}^{(N)} \{s, \xi \}  = {1 \over (2 \pi)^{N} } {\cal P}^{(N,a)} \{s \} 
 \; [1 + O(\e^2)],
\EE
where 
\BE
 {\cal P}^{(N,a)} \{s \} = 
\left(
\prod_{ l {\cal 2 B}_N } 
\oint_{{\cal S}^{1} }
|d \xi_l| \; \right) {\cal P}^{(N)} \{s, \xi \},
\EE
is the $N$-mode {\em amplitude} PDF.
\item
The amplitude variables are almost independent is a sense that for
each $M \ll N$ modes the $M$-mode amplitude PDF is equal to the
product of the one-mode PDF's up to $O(M/N)$ and $o(\e^2)$
corrections,
\BE
{\cal P}^{(M)}_{j_1, j_2, \dots , j_M} = 
 P^{(a)}_{j_1}  P^{(a)}_{j_2} \dots  P^{(a)}_{j_M} \; [1 +
O(M/N) + O(\e^2)].
\EE

\end{enumerate}

\subsection{Why $\psi$'s and not $\phi$'s?}
Importantly, RPA formulation involves independent
{\em phase factors} $\psi = e^\phi$ and not {\em phases}
$\phi$ themselves. Firstly, the phases
would not be convenient because
the mean value of the phases is evolving with the rate equal to
the nonlinear frequency correction \cite{physd}. Thus one could
not say that they are ``distributed uniformly from $-\pi$
to $\pi$''. Moreover the mean fluctuation
of the phase distribution is also growing and they quickly 
spread beyond their initial  $2 \pi$-wide interval \cite{physd}.
But perhaps even more important, $\phi$'s build mutual correlations 
on the nonlinear 
time whereas $\psi$'s remain independent.
Let us give a simple example
illustrating how this property is possible due
to the fact that correspondence between $\phi$
and $\psi$ is not a bijection.
Let $N$ be a
random integer and let $r_1$ and $r_2$ be
two independent (of $N$ and of each other)
random numbers with uniform distribution between $-\pi$ and
$\pi$. Let
$$\phi_{1,2} = 2 \pi N + r_{1,2}.$$
Then 
$$\langle \phi_{1,2}\rangle=  2 \pi \langle N  \rangle,$$
and
$$ \langle \phi_{1}\phi_{2}\rangle=4\pi^2 \langle N^2 \rangle.$$
Thus,
$$
\langle \phi_{1}\phi_{2}\rangle - \langle \phi_{1} \rangle \langle \phi_{2}\rangle
= 4 \pi^2 ( \langle N^2 \rangle - \langle N \rangle^2) > 0,
$$
which means that variables  $\phi_1$ and
$\phi_2$ are correlated.
On the other hand, if we introduce
$$\psi_{1,2} = e^{i \phi_{1,2}},$$
then
$$\langle \psi_{1,2}\rangle=0,$$
and
$$ \langle \psi_{1}\psi_{2}\rangle=0.$$
$$ \langle \psi_{1}\psi_{2}\rangle-
\langle \psi_{1}\rangle  \langle \psi_{2}\rangle =0,$$
which means that variables  $\psi_1$ and
$\psi_2$ are statistically independent.
In this illustrative example it is clear
that the difference in statistical properties 
between $\phi$ and $\psi$ arises from the fact
that function $\psi(\phi)$ does not have inverse
and, consequently, the information about $N$
contained in $\phi$ is lost in $\psi$.

Summarising, statistics of the phase factors $\psi$ is simpler and more
convenient to use than $\phi$ because most of the statistical objects depend only 
on $\psi$. This does not mean, however, that phases  $\phi$ are not observable
and not interesting. Phases $\phi$ can be ``tracked'' in numerical
simulations continuously, i.e. without making jumps to $-\pi$ when the phase
value exceeds $\pi$. Such continuous in $k$ function $\phi(k)$ can achieve a
large range of variation in values due to the dependence of the nonlinear
rotation frequency with $k$. This kind of function implies fastly fluctuating
$\psi(k)$ which is the mechanism behind de-correlation of the
phase factors at different wavenumbers.

\subsection{Wavefields with long spatial correlations.  }

Often in WT, studies are restricted to wavefields with fastly decaying spatial
correlations \cite{Ben}. For such fields, the statistics of the Fourier modes
is close to being Gaussian. Indeed, it the correlation length is much smaller
than the size of the box, then this box can be divided into many smaller
boxes, each larger than the correlation length. The Fourier transform over the
big box will be equal to the sum of the Fourier transforms over the smaller
boxes which are statistically independent quantities. Therefore, by the
Central Limit Theorem, the big-box Fourier transform has a Gaussian
distribution. Corrections to Gaussianity are small as the ratio of the
correlation volume to the box volume. On the other hand, in the RPA defined
above the amplitude PDF is not specified and can significantly deviate from
the Rayleigh distribution (corresponding to Gaussian wavefields). Such fields
correspond to long correlations of order or greater than the box size. In
fact, long correlated fields are quite typical for WT because, due to weak
nonlinearity, wavepackets can propagate over long distance preserving their
identity. Moreover, restricting ourselves to short-correlated fields would
render our study of the PDF evolution meaningless because the later would be
fixed at the Gaussian state. Note that long correlations modify the usual
Wick's rule for the correlator splitting by adding a {\it singular
cumulant}, e.g. for the forth-order correlator,
$$\langle \bar a_j \bar  a_\alpha a_\mu a_\nu\rangle_\psi =
A_j^2 A_\alpha^2 (
\delta^j_\mu \delta^\alpha_\nu +\delta^j_\nu\delta^\alpha_\mu) +
Q_\alpha \delta^\alpha_\nu
\delta^\alpha_\mu
\delta^\alpha_j,
$$
where $Q_\alpha = (A_\alpha^4 - 2A_\alpha^2)$.
These issues were discussed in detail in 
\cite{ln}. 

\subsection{Generating functional.  }

Introduction of generating functionals  simplifies statistical
derivations. It can be defined in several different ways
 to suit a particular
technique. For our problem, the most useful form of the generating
functional is
\BEA
Z^{(N)} \{\lambda, \mu \} =  
{ 1 \over (2 \pi)^{N}} \langle
\prod_{l \in {\cal B}_N }   \; e^{\lambda_l A_l^2} \psi_l^{\mu_l}
\rangle , \label{Z}
\EEA
where $\{\lambda, \mu \} \equiv \{\lambda_l, \mu_l ; l \in {\cal
B}_N\}$ is a set of parameters, $\lambda_l {\cal 2 R}$ and $\mu_l
{\cal 2 Z}$.  

\BE {\cal P}^{(N)} \{s, \xi \} =
 {1 \over (2 \pi)^{N}}  \sum_{\{\mu \}}   \langle
\prod_{l \in {\cal B}_N } \delta (s_l - A_l^2) \, \psi_l^{\mu_l}
\xi_l^{-\mu_l} \rangle \label{pdf-delta} 
\EE
where $\{\mu \} \equiv \{ \mu_l \in {\cal Z}; {l \in {\cal B}_N }
\}$. This expression can be verified by considering mean of a function
$f \{A^2,\psi \}$ using the averaging rule (\ref{pdfn}) and expanding
$f$ in the angular harmonics $\psi_l^m; \; m \in {\cal Z}$ (basis
functions on the unit circle),

\BE f\{A^2,\psi \} = \sum_{\{ m \} } g\{m, A\}
 \, \prod_{l \in {\cal B}_N } \psi_l^{m_l},
\EE
where $\{m\} \equiv \{ m_l \in {\cal Z}; {l \in {\cal B}_N } \}$ are
indices enumerating the angular harmonics. Substituting this into
(\ref{pdfn}) with PDF given by (\ref{pdf-delta}) and taking into
account that any nonzero power of $\xi_l$ will give zero after the
integration over the unit circle, one can see that LHS=RHS, i.e.  that
(\ref{pdf-delta}) is correct. Now we can easily represent
(\ref{pdf-delta}) in terms of the generating functional,
\BE
 {\cal P}^{(N)}  \{s, \xi \} = 
  \hat {\cal L}_\lambda^{-1}
 \sum_{\{\mu \}}
 \left( Z^{(N)} \{\lambda, \mu\}
\, \prod_{l \in {\cal B}_N } \xi_l^{-\mu_l} \right) \label{jointpdf}
\EE
where $\hat {\cal L}_\lambda^{-1}$ stands for inverse the Laplace
transform with respect to all $\lambda_l$ parameters and $\{\mu \}
\equiv \{ \mu_l \in {\cal Z}; {l \in {\cal B}_N } \}$ are the angular
harmonics indices.

Note that we could have defined $Z$ for all real $\mu_l$'s in which
case obtaining $P$ would involve finding the Mellin transform of $Z$
with respect to all $\mu_l$'s.  We will see below however that, given
the random-phased initial conditions, $Z$ will remain zero for all
non-integer $\mu_l$'s.  More generally, the mean of any quantity which
involves a non-integer power of a phase factor will also be
zero. Expression (\ref{jointpdf}) can be viewed as a result of the
Mellin transform for such a special case. It can also be easily
checked by considering the mean of a quantity which involves integer
powers of $\psi_l$'s.

By definition, in RPA fields all variables $A_l$ and $\psi_l$ are
statistically independent and $\psi_l$'s are uniformly distributed on
the unit circle. Such fields imply the following form of the
generating functional
\BE Z^{(N)} \{\lambda, \mu \}  =  Z^{(N,a)}  \{\lambda \}
\, \prod_{l \in {\cal B}_N } \delta(\mu_l),
\label{z-rpa} \EE
where
\BE 
 Z^{(N,a)}  \{\lambda \}
=\langle 
\prod_{l \in {\cal B}_N } 
e^{\lambda_l A_l^2} \rangle
= Z^{(N)} \{\lambda, \mu\}|_{\mu=0}
\EE
is an $N$-mode generating function for the amplitude statistics.
Here, the Kronecker symbol $\delta(\mu_l)$ ensures independence of the
PDF from the phase factors $\psi_l$.  As a first step in validating
the RPA property we will have to prove that the generating functional
remains of form (\ref{z-rpa}) up to $1/N$ and $O(\e^2)$ corrections
over the nonlinear time provided it has this form at $t=0$.

\subsection{One-mode statistics}

Of particular interest are one-mode densities which can
be conveniently obtained using a one-amplitude 
generating function
$$ Z(t,\lambda)=\langle e^{\lambda |a_k|^2}\rangle,
$$
 where $\lambda$ is a real parameter.  Then PDF of the wave
intensities $s= |a_k|^2$ at each ${\bf k}$ can be written as a Laplace
transform,
\begin{equation}
P^{(a)}(s,t) = 
\langle \delta (|a_k|^2 -s) \rangle =
{1 \over 2 \pi}  \int_{0}^{\infty} Z(\lambda, t)e^{-s
\lambda}d\lambda. \label{distribution}
\end{equation}
For the one-point moments of the amplitude we have
\begin{eqnarray} 
M_k^{(p)} \equiv \langle
|a_k|^{2p}\rangle 
=\langle
|a|^{2p}e^{\lambda |a|^2}\rangle|_{\lambda=0}=\nonumber\\Z_{\lambda \cdots \lambda}|_{\lambda=0}
= \int_{0}^{\infty} s^p P^{(a)}(s, t) \, ds \label{momPDF}, 
\end{eqnarray}
where $p \in {\cal N}$ and subscript $\lambda$ means differentiation
with respect to $\lambda$ $p$ times.

The first of these moments, $n_k = M_k^{(1)}$, is the waveaction spectrum.
Higher moments $M_k^{(p)}$ measure fluctuations of the waveaction $k$-space
distributions about their
mean values \cite{ln}. In particular the r.m.s. value of these fluctuations is 
\BE
\sigma_k = \sqrt{ M_k^{(2)} - n_k^2}.
\EE
 
\section{Separation of timescales: general idea}

When the wave amplitudes are small, the nonlinearity is weak and the wave
periods, determined by the linear dynamics, are much smaller than the
characteristic time at which different wave modes exchange energy. In the
other words, weak nonlinearity results in a timescale separation and our goal
will be to describe the slowly changing wave statistics by averaging over the
fast linear oscillations.
To filter out fast oscillations, we will seek
seek for the solution at time $T$ such that $2 \pi / \omega \ll T \ll
\tau_{nl}$. Here $\tau_{nl}$ is the characteristic time of nonlinear
evolution which, as we will see later is $ \sim
1/\omega \epsilon^2$ for the three-wave systems and
$ \sim 1/\omega \epsilon^4$ for the four-wave systems.
   Solution at $t=T$ can be sought at series in small
 small nonlinearity parameter $\epsilon$,
\BE
a_l(T)=a_l^{(0)}+\epsilon a_l^{(1)}+\epsilon^2 a_l^{(2)}.
\label{Expansion}
\EE
Then we are going to iterate the equation of motion
(\ref{Interaction}) or  (\ref{FourWaveEquationOfMotionA}) 
to obtain $a_l^{(0)}$,
$a_l^{(1)}$ and $a_l^{(2)}$ by iterations. 


During this analysis the certain integrals of a type 
$$f(t) = \int\limits_0^t g(t) \exp(i \omega t)$$
will play a crucial role. Following  \cite{Ben} we introduce

\BEA
\Delta^l_{mn}=\int_0^T e^{i\omega^l_{mn}t}d t =
\frac{({e^{i\omega^l_{mn}T}-1})}{{i \omega^l_{mn}}},  \nonumber\\
\Delta^{kl}_{mn}=\int_0^T e^{i\omega^{kl}_{mn}t}d t =
\frac{({e^{i\omega^{kl}_{mn}T}-1})}{{i \omega^{kl}_{mn}}},  \nonumber\\
\label{NewellsDelta}
\EEA
and
$$E(x,y)=\int_0^T \Delta(x-y)e^{i y t} d t .$$
We will be interested in  a long time asymptotics of the above expressions, 
so the following properties will be useful:
$$\lim\limits_{T\to\infty}E(0,x)= T (\pi\delta(x)+iP(\frac{1}{x})),$$
and
$$
\lim\limits_{T\to\infty}|\Delta(x)|^2=2\pi T\delta(x).$$

\section{Weak nonlinearity expansion: three-wave case.}
Substituting the expansion (\ref{Expansion}) in (\ref{Interaction}) we
get in the zeroth order
$$ a_l^{(0)}(T)=a_l(0)\label{definitionofa} ,$$
i.e. the zeroth order term is time independent. This corresponds to
the fact that the interaction representation wave amplitudes are
constant in the linear approximation.  For simplicity, we will write
$a^{(0)}_l(0)= a_l$, understanding that a quantity is taken at $T=0$
if its time argument is not mentioned explicitly.  

Here we have taken into account that $a^{(0)}_l(T)= a_l$ and
$a^{(1)}_k (0)=0$.  

The first order is
given by
\BEA
a^{(1)}_l (T) = -i \sum_{m,n=1}^\infty \left(   V^l_{mn}
a_m a_n \Delta^l_{mn} \delta^l_{m+n}\right.\CR\left.\hskip 4cm
+
2 \bar{V}^m_{ln}a_m\bar{a}_n \bar\Delta^m_{ln}\delta^m_{l+n}
\right),
\label{FirstIterate}
\EEA
Here we have taken into account that $a^{(0)}_l(T)= a_l$ and
$a^{(1)}_k (0)=0$.  Perform the second iteration, and integrate over
time to obtain
To calculate the second iterate, write

\BEA i\dot{ a}^{(2)}_l  = \sum_{m,n=1}^\infty \Big[   V^l_{mn}\delta^l_{m+n} e^{i \omega^l_{mn} t}\left(a_m^{(0)} a_n^{(1)}+ a_m^{(1)} a_n^{(0)}\right)
\CR \hspace{3cm}+2\bar{V}^m_{ln}\delta^m_{l+n} e^{ -i \omega^m_{ln} t}\left(a_m^{(1)} \bar{a}_n^{(0)}+ a_m^{(0)} \bar{a}_n^{(1)}\right)\Big].\CR\label{SecondIterateTimeDerivative}
\EEA 

Substitute (\ref{FirstIterate}) into (\ref{SecondIterateTimeDerivative}) and integrate over time to obtain
\BEA a_l^{(2)} (T)  &=& \sum_{m,n, \mu, \nu=1}^\infty \left[ 2 V^l_{mn} 
\left(
-V^m_{\mu \nu}a_n a_\mu a_\nu E[\omega^l_{n \mu \nu},\omega^l_{mn}]
\delta^m_{\mu + \nu} 
-2 
\bar V^\mu_{m \nu}a_n a_\mu \bar a_\nu \bar 
E[\omega^{l \nu}_{n \mu},\omega^l_{mn}]\delta^\mu_{m + \nu}\right)
\delta^l_{m+n} \right.\CR && \left.
+ 2 
\bar V^m_{ln} 
 \left(
-V^m_{\mu \nu}\bar a_n a_\mu a_\nu E[\omega^{ln}_{\mu \nu},-\omega^m_{ln}]
\delta^m_{\mu + \nu} 
-
2 \bar V^\mu_{m \nu}\bar a_n a_\mu \bar a_\nu 
E[-\omega^\mu_{n \nu l},-\omega^m_{l n}]  \delta^\mu_{m + \nu}
\right) \delta^m_{l+ n} \right. \CR && \left.
+ 2 
\bar V^m_{ln} 
 \left(
\bar V^n_{\mu \nu}a_m \bar a_\mu  \bar a_\nu \delta^n_{\mu + \nu}
E[-\omega^m_{l\nu\mu},-\omega^m_{ln}] 
+
2  V^\mu_{n \nu}a_m \bar a_\mu  a_\nu E[\omega^{\mu l}_{\nu m},
-\omega^m_{ln}]\delta^\mu_{n + \nu}\right)\delta^m_{l+n}
\right],\CR\label{SecondIterate}
\EEA
\noindent
where we used $a^{(2)}_k (0)=0$. Hereafter, we drop super-script $ (0)$ in
expressions like  (\ref{SecondIterate}) for brevity of notations.

\section{Weak nonlinearity expansion: four-wave case}

Substituting (\ref{Expansion}) in (\ref{FourWaveEquationOfMotionA}) we
get in the zeroth order
$ a_l^{(0)}(T)=a_l(0)\label{definitionofa2} $,
and the first
iteration of (\ref{FourWaveEquationOfMotionA}) gives
\begin{eqnarray}
a_l\one(T) = - i \sum_{\alpha\mu\nu} W^{l\alpha}_{\mu\nu} \bar
 a_\alpha a_\mu a_\nu \delta^{l\alpha}_{\mu\nu} \Delta^{l
 \alpha}_{\mu\nu} + i \Omega_l a_l T.  \label{FirstIterate1}
\end{eqnarray}
Iterating one more time we get
\begin{eqnarray}
a_l\two(T)&=&\sum_{\alpha\mu\nu v u}\left( 
W^{\mu\nu}_{\alpha u} W^{lu}_{v\beta} \delta^{\mu\nu}_{\alpha u} \delta^{lu}_{v\beta}
 a_{\alpha} a_v a_\beta 
\bar a_{\mu} \bar a_{\nu} 
E(\tilde\omega^{l\mu\nu}_{\alpha v \beta}, \tilde\omega^{lu}_{v\beta}) 
\right.
\nonumber \\
&&
\left.
- 
2 W^{\alpha v}_{\mu\nu} W^{l u}_{v \beta}
\delta^{\alpha v}_{\mu\nu} \delta^{l u}_{v \beta}
\bar a_{\alpha} \bar a_{u}
a_{\mu} a_{\nu} a_\beta 
E(\tilde \omega^{l\alpha u}_{\mu\nu\beta}, \tilde \omega^{lu}_{v\beta})\right) 
-\Omega_l^2 a_l \frac{T^2}{2}
\nonumber  \\ 
&&+ 
\sum_{\alpha\mu\nu}\left(
\Omega_l 
W^{l\alpha}_{\mu\nu}
\delta^{l\alpha}_{\mu\nu}
\bar a_{\alpha} a_{\mu} a_{\nu} 
E(\tilde\omega^{l\alpha}_{\mu\nu},0)  
-W^{l \alpha}_{\mu\nu }
\delta^{l \alpha}_{\mu\nu }\bar{ a_{\alpha}} 
a_{\mu} a_{\nu}(\Omega_\alpha
-2\Omega_\nu) 
\int \limits_0^T
\tau e^{i\Omega^{l \alpha}_{\mu\nu } \tau} d \tau
\right). 
\label{SecondIterate1}
\end{eqnarray}
%

\section{Asymptotic expansion of the generating functional.}

%
%
Let us first obtain an asymptotic weak-nonlinearity expansion
for the generating functional
$Z\{\lambda, \mu\}$ exploiting the separation of the linear and
nonlinear time scales. \footnote{ Hereafter we omit superscript ${(N)}$
in the $N$-mode objects if it does not lead to a confusion.}  To
do this, we have to calculate $Z$ at the intermediate time $t=T$ via
substituting into it $a_j(T)$ from  (\ref{Expansion})
For the amplitude and phase ``ingredients'' in $Z$ we have,
\BE e^{\lambda_j |a_j|^2}= e^{\lambda_j |a_j\zer +\e a_j\one
+\e^2 a_j\two|^2} =
e^{\lambda_j A_j^{(0) 2}} ( 1+\e {\alpha_{1j}} + \e^2
{\alpha_{2j}}),
\label{alpha}\EE
and
\BE \psi_j^{\mu_j}=\left(\frac{a_j\zer +\e a_j\one +\e^2
a_j\two}{\bar a_j\zer +\e \bar a_j\one +\e^2 \bar
a_j\two}\right)^{\frac{\mu_j}{2}}=
= \psi_j^{(0)\mu_j} (1+\e {\beta_{1j}}
+\e^2 {\beta_{2j}} ),
\label{beta}\EE
where
\BEA
{\alpha_{1j}} &=& \lambda_j(a_j\one\bar a_j\zer + \bar a_j\one
a_j\zer),  \\
 {\alpha_{2j}} &=&  {(\lambda_j +\lambda_j^2
A_j^{(0)2}|a_j\one|^2 +\lambda_j(a_j\two \bar a_j\zer + \bar a_j\two
a_j\zer)+\frac{\lambda_j^2}{2}(a_j\one \bar a_j\zer)^2 + (\bar
a_j\one a_j\zer)^2}, \\
{\beta_{1j}} &=&
\frac{\mu_j}{2A_j^{(0)2}}(a_j\one\bar a_j\zer-\bar a_j\one
a_j\zer), \\
{\beta_{2j}} &=& \frac{\mu_j}{2A_j^{(0)2}}(a_j\two\bar
a_j\zer-\bar a_j\two a_j\zer)+\frac{\mu_j}{4}\left[
\left(\frac{\mu_j}{2}-1\right)\left(\frac{a_j\one}{a_j\zer}\right)^2
+\left(\frac{\mu_j}{2}+1\right)\left(\frac{\bar a_j\one}{\bar
a_j\zer}\right)^2
\right]-\frac{\mu_j^2|a_j\one|^2}{4A_j^{(0)2}}.
\EEA
Substituting expansions (\ref{alpha}) and (\ref{beta}) into the
expression for $Z$, we have 
\BE
 Z\{\lambda, \mu, T\} =  X\{\lambda, \mu,T\} +  \bar X \{\lambda, - \mu,T\} 
\label{zx}
\EE
with 
\BE
X\{\lambda, \mu,T\} =  X\{\lambda, \mu,0\} +  (2 \pi)^{2N} \left<\prod_{\|l\|<N}
e^{\lambda_l|a_l\zer|^2}[\e J_1 +\e^2(J_2 +J_3+J_4+J_5)] \right>_A + O(\e^4), 
\label{xt}
\EE
where
\BEA
J_1 &=&   \left<\prod_l \psi_l^{(0)\mu_l} 
\sum_j (\lambda_j
+\frac{\mu_j}{2|a_j\zer|^2})a_j\one\bar a_j\zer \right>_\psi, \label{j1} \\
J_2 &=&   {1 \over 2} \left<\prod_l \psi_l^{(0)\mu_l} 
\sum_j (\lambda_j+
\lambda_j^2|a_j\zer|^2-\frac{\mu_j^2}{2|a_j\zer|^2})|a_j\one|^2
 \right>_\psi, \label{j2} \\
J_3 &=&   \left<\prod_l \psi_l^{(0)\mu_l} 
 \sum_j 
(\lambda_j + \frac{\mu_j}{2|a_j\zer|^2})a_j\two\bar a_j\zer 
 \right>_\psi, \label{j3} \\
J_4 &=&   \left<\prod_l \psi_l^{(0)\mu_l} 
\sum_j 
\left[\frac{\lambda_j^2}{2}+\frac{\mu_j}{4|a_j\zer|^4}\left(\frac{\mu_j}{2}-1\right)+\frac{\lambda_j
\mu_j}{2|a_j\zer|^2} \right](a_j\one \bar a_j\zer)^2
 \right>_\psi, \label{j4} \\
J_5 &=&   {1 \over 2} \left<\prod_l \psi_l^{(0)\mu_l} 
\sum_{j \ne k}\lambda_j\lambda_k(a_j\one\bar a_j\zer +\bar a_j\one
a_j\zer)a_k\one\bar a_k\zer + (\lambda_j
+\frac{\mu_j}{4|a_j\zer|^2})\frac{\mu_k}{|a_k\zer|^2}(a_k\one\bar a_k\zer
-\bar a_k\one a_k\zer)a_j\one\bar a_j\zer
 \right>_\psi, \label{j5} 
\EEA
where $\left< \cdot \right>_A$ and $\left< \cdot \right>_\psi$ denote
the averaging over the initial amplitudes and initial phases
(which can be done independently).
Note that so far our calculation for $Z(T)$ is the same for the three-wave and
for the four-wave cases.   Now we have to substitute expressions for $a\one$
and $a\two$ which are different for the three-wave and the four-wave cases
and given by 
 (\ref{FirstIterate}), (\ref{SecondIterate}) and  (\ref{FirstIterate}),
(\ref{SecondIterate})
 respectively.

\section{Evolution of statistics of three-wave systems}

\subsection{Equation for the generating functional}

Let us consider the initial fields $a_k(0) = a^{(0)}_k$ 
which are of the RPA type as defined above. We will perform
averaging over the statistics of the initial fields
in order to obtain an evolution equations, first for $Z$ and
then for the multi-mode PDF. 
Let us  introduce a graphical  classification of the above terms which
will allow us to simplify the statistical averaging and to understand
which terms are dominant. We will only consider here  contributions from
 $J_1$ and  $J_2$ which will allow us to understand the basic method.
Calculation of the rest of the terms, $J_3$, $J_4$ and  $J_5$, 
follows the same principles and can be found in \cite{physd}.
First, 
 The linear in $\e$ terms are represented by $J_1$ which, upon using
(\ref{FirstIterate}), becomes
\BEA 
J_1
&=&\left<\prod_l \psi_l^{\mu_l}
\sum_{j,m,n}(\lambda_j +\frac{\mu_j}{2A_j^2})
\left(V_{mn}^j
a_m a_n\Delta_{mn}^j\delta_{m+n}^j +2\bar V_{jn}^m a_m\bar a_n
\bar\Delta_{jn}^m\delta_{j+n}^m\right)\bar
a_j\right>_\psi.
\label{firstineps}\EEA
Let us introduce some graphical notations for a simple classification
of different contributions to this and to other (more lengthy)
formulae that will follow. Combination $V_{mn}^j \delta_{m+n}^j$ will
be marked by a vertex joining three lines with in-coming $j$ and
out-coming $m$ and $n$ directions.  Complex conjugate $\bar V_{mn}^j
\delta_{m+n}^j$ will be drawn by the same vertex but with the opposite
in-coming and out-coming directions.  Presence of $a_j$ and $\bar a_j$
will be indicated by dashed lines pointing away and toward the vertex
respectively.~\footnote{This technique provides a useful
classification method but not a complete mathematical description of
the terms involved.} Thus, the two terms in formula (\ref{firstineps})
can be schematically represented as follows,

\

\vskip 1cm

\
\BEA
C_1=
\parbox{40mm} {
\begin{fmffile}{one}
   \begin{fmfgraph*}(110,62)
    \fmfleft{i1,i2}
    \fmfright{o1}
    \fmflabel{$m$}{i1}
    \fmflabel{$n$}{i2}
    \fmflabel{$j$}{o1}
    \fmf{dashes_arrow}{v1,i1}
    \fmf{dashes_arrow}{v1,i2}
 \fmf{dashes_arrow}{o1,v1}
   \end{fmfgraph*}
\end{fmffile}
}
&\quad  and \quad&
C_2=
\parbox{40mm} {
\begin{fmffile}{two}
   \begin{fmfgraph*}(110,62)
    \fmfleft{i1,i2}
    \fmfright{o1}
    \fmflabel{$m$}{i1}
    \fmflabel{$n$}{i2}
    \fmflabel{$j$}{o1}
    \fmf{dashes_arrow}{v1,i1}
    \fmf{dashes_arrow}{i2,v1}
 \fmf{dashes_arrow}{o1,v1}
   \end{fmfgraph*}
\end{fmffile}
}
\nonumber
\EEA
\vskip 1cm
Let us average over all the independent phase factors in the set
$\{\psi\}$.
Such averaging takes into account the statistical
independence and uniform distribution of
variables $\psi$. In particular, $\langle\psi\rangle=0$,
$\langle\psi_l \psi_m\rangle=0$ and $\langle\psi_l \bar
\psi_m\rangle=\delta^m_l$. Further, the products that involve
odd number of $\psi$'s are always zero, and among the even
products only those can survive that have equal numbers of
$\psi$'s and $\bar \psi$'s. These $\psi$'s and $\bar \psi$'s
must cancel each other which is possible if their indices
are matched in a pairwise way similarly to the  Wick's
theorem. The difference with the standard Wick, however, is
that there exists possibility of not only internal
(with respect to the sum) matchings but also external
ones with $\psi$'s in the pre-factor $\Pi \psi_l^{\mu_l}$.

Obviously, non-zero contributions can only arise for terms in which
all $\psi$'s cancel out either via internal mutual couplings within
the sum or via their external couplings to the $\psi$'s in the
$l$-product.  The internal couplings will indicate by joining the
dashed lines into loops whereas the external matching will be shown as
a dashed line pinned by a blob at the end. The number of blobs in
a particular graph will be called the {\em valence} of this graph.

Note that there will be no
contribution from the internal couplings between the incoming and the
out-coming lines of the same vertex because, due to the
$\delta$-symbol, one of the wavenumbers is 0 in this case, which means
\footnote{In the present paper we consider only spatially homogeneous wave
turbulence fields. In spatially homogeneous fields, due to momentum
conservation, there is no coupling to the zero mode $k=0$ because such
coupling would violate momentum conservation. Therefore if one of the
arguments of the interaction matrix element $V$ is equal to zero, the
matrix element is identically zero. That is to say that for any
spatially homogeneous wave turbulence system
$ V^{k=0}_{k_1 k_2} =  V^{k}_{k_1=0 k_2}  = V^{k}_{k_1 k_2=0} =0.$}
that $V=0$. For $J_1$ we have
$$ J_1 = \langle C_1 \rangle_\psi + \langle C_2 \rangle_\psi,$$ with
\BEA
 \langle C_1 \rangle_\psi  =
\parbox{40mm} {
\begin{fmffile}{three}
   \begin{fmfgraph*}(110,62)
    \fmfleft{i1,i2}
    \fmfright{o1}
    \fmflabel{$m$}{i1}
    \fmflabel{$n$}{i2}
    \fmflabel{$j$}{o1}
    \fmf{dashes_arrow}{v1,i1}
    \fmf{dashes_arrow}{v1,i2}
 \fmf{dashes_arrow}{o1,v1}
\fmfdot{i1}
\fmfdot{i2}
\fmfdot{o1}
   \end{fmfgraph*}
\end{fmffile}
}
&\quad  + \quad \quad&
\parbox{40mm} {
\begin{fmffile}{four}
   \begin{fmfgraph*}(110,62)
    \fmfleft{i1}
   \fmfforce{0.5w,0.5h}{v1}
    \fmfright{o1}
    \fmflabel{$m$}{i1}
    \fmflabel{$2m$}{o1}
    \fmf{dashes_arrow, left=.4, tension=.3}{v1,i1}
    \fmf{dashes_arrow, right=.4, tension=.3 }{v1,i1}
 \fmf{dashes_arrow}{o1,v1}
\fmfdot{i1}
\fmfdot{o1}
   \end{fmfgraph*}
\end{fmffile}
}
\nonumber
\EEA
and
\BEA
 \langle C_2 \rangle_\psi  =
\parbox{40mm} {
\begin{fmffile}{five}
   \begin{fmfgraph*}(110,62)
    \fmfleft{i1,i2}
    \fmfright{o1}
    \fmflabel{$m$}{i1}
    \fmflabel{$n$}{i2}
    \fmflabel{$j$}{o1}
    \fmf{dashes_arrow}{v1,i1}
    \fmf{dashes_arrow}{i2,v1}
 \fmf{dashes_arrow}{o1,v1}
\fmfdot{i1}
\fmfdot{i2}
\fmfdot{o1}
   \end{fmfgraph*}
\end{fmffile}
}
&\quad  + \quad \quad&
\parbox{40mm} {
\begin{fmffile}{six}
   \begin{fmfgraph*}(110,62)
    \fmfleft{i1}
   \fmfforce{0.5w,0.5h}{v1}
    \fmfright{o1}
    \fmflabel{$n$}{i1}
    \fmflabel{$2n$}{o1}
    \fmf{dashes_arrow, left=.4, tension=.3}{i1,v1}
    \fmf{dashes_arrow, right=.4, tension=.3 }{i1,v1}
 \fmf{dashes_arrow}{v1,o1}
\fmfdot{o1}
\fmfdot{i1}
   \end{fmfgraph*}
\end{fmffile}
}
\nonumber
\EEA
\vskip 1cm
which correspond to the following expressions,
\vskip 1cm
\BEA 
 \langle C_1 \rangle_\psi  
&=&
\sum_{j \ne m \ne n}(\lambda_j +\frac{\mu_j}{2A_j^2})
V_{mn}^j
A_m A_n A_j \Delta_{mn}^j\delta_{m+n}^j \delta(\mu_m +1)
\delta(\mu_n +1) \delta(\mu_j-1) 
 \prod_{l \ne j,m,n} \delta (\mu_l)
 \nonumber \\
&+& \sum_{m}(\lambda_{2m} +\frac{\mu_{2m}}{2A_{2m}^2})
V_{mm}^{2m}
A_m^2  A_{2m} \Delta_{mm}^{2m} \delta(\mu_m +2)
 \delta(\mu_{2m}-1) 
 \prod_{l \ne m,2m} \delta (\mu_l)
\label{haha1}
\EEA
and
\BEA 
 \langle C_2 \rangle_\psi  =
&=&
2\sum_{j \ne m \ne n}(\lambda_j +\frac{\mu_j}{2A_j^2})
\bar V_{jn}^m A_m A_n A_j
\bar\Delta_{jn}^m \delta_{j+n}^m
 \delta(\mu_m +1) \delta(\mu_n -1) \delta(\mu_j-1) 
 \prod_{l \ne j,m,n} \delta (\mu_l)
 \nonumber \\
&+&
2\sum_{n}(\lambda_n +\frac{\mu_n}{2A_n^2})
\bar V_{nn}^{2n} A_{2n} A_n^2
\bar\Delta_{nn}^{2n}
 \delta(\mu_{2n} +1) \delta(\mu_n -2) 
 \prod_{l \ne n,2n} \delta (\mu_l).
\label{haha2}\EEA
Because  of  the $\delta$-symbols  involving  $\mu$'s,  it takes  very
special combinations  of the arguments $\mu$  in $Z\{ \mu  \}$ for the
terms  in  the above  expressions  to  be  non-zero.  For  example,  a
particular term in  the first sum of (\ref{haha1})  may be non-zero if
two $\mu$'s in the set $\{ \mu  \}$ are equal to 1 whereas the rest of
them are 0. But in this case  there is only one other term in this sum
(corresponding to the  exchange of values of $n$ and  $j$) that may be
non-zero too.   In fact,  only utmost  two terms in  the both
(\ref{haha1})  and (\ref{haha2}) can  be non-zero  simultaneously.  In
the  other words,  each external  pinning of  the dashed  line removes
summation in  one index and, since  all the indices are  pinned in the
above diagrams, we are left with no summation at all in $J_1$ i.e. the
number of terms in $J_1$ is $O(1)$ with respect to large $N$.  We will
see   later   that    the   dominant   contributions   have   $O(N^2)$
terms. Although  these terms  come in the  $\e^2$ order, they  will be
much greater  that the $\e^1$ terms  because the limit  $N \to \infty$
must always be taken before $\e \to 0$.

Let us consider the first of the $\e^2$-terms, $J_2$. Substituting
 (\ref{FirstIterate}) into (\ref{j2}), we have
\vskip 1cm
\BEA J_2 &=& \frac{1}{2}\langle\prod_l\psi_l^{(0)\mu_l}
\sum_{j,m,n,\kappa,\nu}(\lambda_j+\lambda_j^2A_j^2-\frac{\mu_j^2}{2A_j^2})\CR 
&&\hspace{1cm} \times (V_{mn}^ja_ma_n\Delta_{mn}^j\delta_{m+n}^j+2\bar V_{jn}^ma_m\bar a_n\bar\Delta_{jn}^m\delta_{j+n}^m)
(\bar V_{\kappa\nu}^j\bar a_{\kappa}\bar a_{\nu}\bar\Delta_{\kappa\nu}^j\delta_{\kappa +\nu}^j
+2V_{j\nu}^{\kappa}\bar a_{\kappa}a_{\nu}\Delta_{j\nu}^{\kappa}\delta_{j+\nu}^{\kappa})\rangle_{\psi}\CR 
&=& \langle B_1 + B_2 + \bar B_2 + B_3 \rangle_{\psi}, \EEA
where
\\
\BEA
B_1 = \hspace{1cm}
\parbox{35mm} {
\begin{fmffile}{n7}
   \begin{fmfgraph*}(60,45)
\fmfforce{(0.w,1.h)}{i1}
\fmfforce{(0.w,0.h)}{i2}
\fmfforce{(1.w,1.h)}{o1}
\fmfforce{(1.w,0.h)}{o2}
\fmfforce{(0.25w,0.5h)}{v1}
\fmfforce{(0.75w,0.5h)}{v2}
\fmfforce{(0.5w,1.h)}{v3}
    \fmflabel{$m$}{i1}
    \fmflabel{$n$}{i2}
    \fmflabel{$\kappa$}{o1}
    \fmflabel{$\nu$}{o2}
    \fmf{dashes_arrow}{v1,i1}
    \fmf{dashes_arrow}{v1,i2}
    \fmf{dashes_arrow}{o1,v2}
    \fmf{dashes_arrow}{o2,v2}
\fmf{dots_arrow, label=$j$}{v2,v1}
   \end{fmfgraph*}
\end{fmffile}
}
&
B_2 = \hspace{1cm}
\parbox{35mm} {
\begin{fmffile}{n8}
   \begin{fmfgraph*}(60,45)
\fmfforce{(0.w,1.h)}{i1}
\fmfforce{(0.w,0.h)}{i2}
\fmfforce{(1.w,1.h)}{o1}
\fmfforce{(1.w,0.h)}{o2}
\fmfforce{(0.25w,0.5h)}{v1}
\fmfforce{(0.75w,0.5h)}{v2}
\fmfforce{(0.5w,1.h)}{v3}
    \fmflabel{$m$}{i1}
    \fmflabel{$n$}{i2}
    \fmflabel{$\kappa$}{o1}
    \fmflabel{$\nu$}{o2}
    \fmf{dashes_arrow}{v1,i1}
    \fmf{dashes_arrow}{v1,i2}
    \fmf{dashes_arrow}{o1,v2}
    \fmf{dashes_arrow}{v2,o2}
\fmf{dots_arrow, label=$j$}{v2,v1}
   \end{fmfgraph*}
\end{fmffile}
} and
&
B_3 = \hspace{1cm}
\parbox{35mm} {
\begin{fmffile}{n9}
   \begin{fmfgraph*}(60,45)
\fmfforce{(0.w,1.h)}{i1}
\fmfforce{(0.w,0.h)}{i2}
\fmfforce{(1.w,1.h)}{o1}
\fmfforce{(1.w,0.h)}{o2}
\fmfforce{(0.25w,0.5h)}{v1}
\fmfforce{(0.75w,0.5h)}{v2}
\fmfforce{(0.5w,1.h)}{v3}
    \fmflabel{$m$}{i1}
    \fmflabel{$n$}{i2}
    \fmflabel{$\kappa$}{o1}
    \fmflabel{$\nu$}{o2}
    \fmf{dashes_arrow}{v1,i1}
    \fmf{dashes_arrow}{i2,v1}
    \fmf{dashes_arrow}{o1,v2}
    \fmf{dashes_arrow}{v2,o2}
\fmf{dots_arrow, label=$j$}{v2,v1}
   \end{fmfgraph*}
\end{fmffile}
}
\EEA 
\\
\\
\vskip 1cm
Here the graphical notation for the interaction coefficients $V$ and 
the amplitude $a$ is the same as introduced in the previous section and
the dotted line with index $j$ indicates that there is a summation over $j$ 
but there is no amplitude $a_j$ in the corresponding expression.

Let us now perform the phase averaging which corresponds to the internal and external
couplings of the dashed lines. For $\langle B_1\rangle_{\psi}$ we have
\vskip 1cm
\BEA
\langle B_1\rangle_{\psi} =  \hspace{1cm}
\parbox{35mm} {
\begin{fmffile}{n10}
   \begin{fmfgraph*}(60,50) \fmfkeep{x}
\fmfforce{(0.w,1.h)}{i1}
\fmfforce{(0.w,0.h)}{i2}
\fmfforce{(1.w,1.h)}{o1}
\fmfforce{(1.w,0.h)}{o2}
\fmfforce{(0.25w,0.5h)}{v1}
\fmfforce{(0.75w,0.5h)}{v2}
\fmfforce{(0.5w,1.h)}{v3}
    \fmflabel{$m$}{i1}
    \fmflabel{$n$}{i2}
    \fmflabel{$\kappa$}{o1}
    \fmflabel{$\nu$}{o2}
    \fmf{dashes_arrow}{v1,i1}
    \fmf{dashes_arrow}{v1,i2}
    \fmf{dashes_arrow}{o1,v2}
    \fmf{dashes_arrow}{o2,v2}
\fmf{dots_arrow, label=$j$}{v2,v1}
\fmfdot{i1}
\fmfdot{i2}
\fmfdot{o1}
\fmfdot{o2}
   \end{fmfgraph*}
\end{fmffile}
} 
+
&
\parbox{35mm} {
\begin{fmffile}{n11}
   \begin{fmfgraph*}(70,50) \fmfkeep{fish}
\fmfforce{(0.w,1.h)}{i1}
\fmfforce{(0.w,0.h)}{i2}
\fmfforce{(1.w,.5h)}{o1}
\fmfforce{(0.2w,0.5h)}{v1}
\fmfforce{(0.6w,0.5h)}{v2}
    \fmflabel{$m$}{i1}
    \fmflabel{$n$}{i2}
    \fmf{dashes_arrow}{v1,i1}
    \fmf{dashes_arrow}{v1,i2}
    \fmf{dashes_arrow, left=1., label= $\nu$}{o1,v2}
    \fmf{dashes_arrow, right=1., label= $\nu$}{o1,v2}
\fmf{dots_arrow, label=$2 \nu$}{v2,v1}
\fmfdot{i1}
\fmfdot{i2}
\fmfdot{o1}
   \end{fmfgraph*}
\end{fmffile}
} 
+
&
\parbox{35mm} {
\begin{fmffile}{n12}
   \begin{fmfgraph*}(70,50) 
\fmfforce{(0.w,1.h)}{i1}
\fmfforce{(0.w,0.h)}{i2}
\fmfforce{(1.w,.5h)}{o1}
\fmfforce{(0.2w,0.5h)}{v1}
\fmfforce{(0.6w,0.5h)}{v2}
    \fmflabel{$\kappa$}{i1}
    \fmflabel{$\nu$}{i2}
    \fmf{dashes_arrow}{i1,v1}
    \fmf{dashes_arrow}{i2,v1}
    \fmf{dashes_arrow, left=1., label= $m$}{v2,o1}
    \fmf{dashes_arrow, right=1., label= $m$}{v2,o1}
\fmf{dots_arrow, label=$2 m$}{v1,v2}
\fmfdot{i1}
\fmfdot{i2}
\fmfdot{o1}
   \end{fmfgraph*}
\end{fmffile}
} \nonumber
\EEA
\BEA 
+2 \hspace{.5cm} 
\parbox{35mm} {
\begin{fmffile}{n13}
   \begin{fmfgraph*}(70,50) \fmfkeep{theta}
\fmfforce{(0.1w,0.5h)}{v1}
\fmfforce{(0.9w,0.5h)}{v2}
\fmf{dots_arrow, label=$j$}{v2,v1}
    \fmf{dashes_arrow, left=.7, label= $m$}{v1,v2}
    \fmf{dashes_arrow, right=.7, label= $n$}{v1,v2}
   \end{fmfgraph*}
\end{fmffile}
} ,
\label{B_1diagram}
\EEA
\\
where
\\
\BEA 
\parbox{25mm} {\fmfreuse{x}}
&=&
\frac{1}{2}\sum_{j\neq m\neq n\neq\kappa\neq\nu}(\lambda_j+\lambda_j^2A_j^2-\frac{\mu_j^2}{2A_j^2})
V_{mn}^j\bar V_{\kappa\nu}^j\Delta_{mn}^j\bar\Delta_{\kappa\nu}^j\delta_{m+n}^j\delta_{\kappa+\nu}^jA_mA_nA_{\kappa}A_{\nu}
\nonumber \\
&&  \hspace{2.2cm}  \times
\delta(\mu_m+1)\delta(\mu_n+1)\delta(\mu_{\kappa}-1)\delta(\mu_{\nu}-1)
\prod_{l\neq m,n,\kappa,\nu}\delta(\mu_l)
\nonumber
\EEA
\\
\\
\BEA 
\parbox{35mm} {\fmfreuse{fish}}
&=&\frac{1}{2}\sum_{m\neq n\neq\nu}(\lambda_{2\nu}+\lambda_{2\nu}^2A_{2\nu}^2-\frac{\mu_{2\nu}^2}{2A_{2\nu}^2})
V_{mn}^{2\nu}\delta_{m+n}^{2\nu}\bar V_{\kappa\nu}^{2\nu}
\nonumber\\ 
 &&\times
\Delta_{mn}^{2\nu}\bar\Delta_{\kappa\nu}^{2\nu}A_mA_nA_{\nu}^2
\delta(\mu_m+1)\delta(\mu_n+1)\delta(\mu_{\nu}-2)
\prod_{l\neq m,n,\nu}\delta(\mu_l)
\nonumber
\EEA
\\
\BEA 
2 \hspace{.5cm} \parbox{35mm} {\fmfreuse{theta}}
=\prod_{l}\delta(\mu_l)\sum_{j,m,n}(\lambda_{j}+\lambda_{j}^2A_{j}^2
)
|V_{mn}^{j}|^2
|\Delta_{mn}^{j}|^2
\delta_{m+n}^{j}A_m^2A_n^2
\nonumber
\EEA
\\
\\
\vskip 1cm
We have not written out the third term in (\ref{B_1diagram}) because
it is just a complex conjugate of the second one. Observe that all the
diagrams in the first line of (\ref{B_1diagram}) are $O(1)$ with
respect to large $N$ because all of the summations are lost due to the
external couplings(compare with the previous section). On the other
hand, the diagram in the second line contains two purely-internal
couplings and is therefore $O(N^2)$. This is because the number of
indices over which the summation survives is equal to the number of
purely internal couplings. Thus, the zero-valent graphs
are dominant and we can write
\BEA \langle B_1\rangle_\psi = \prod_l\delta(\mu_l)\sum_{j,m,n}(\lambda_{j}+\lambda_{j}^2A_{j}^2
)
|V_{mn}^{j}|^2
|\Delta_{mn}^{j}|^2
\delta_{m+n}^{j}A_m^2A_n^2[1+O(1/N^2)]\EEA

Now we are prepared to understand a general rule: the dominant contribution
always comes from the graphs with minimal valence, because each external
pinning reduces the number of summations by one. The minimal valence of
the graphs in  $\langle B_2\rangle_\psi$ is one and, therefore, 
 $\langle B_2\rangle_\psi$ is order $N$ times smaller than  $\langle B_1\rangle_\psi$.
On the other hand, $\langle B_3\rangle_\psi$ is of the same order as
 $\langle B_1\rangle_\psi$, because it contains zero-valence graphs. We have
%
%
\BEA
\langle B_3 \rangle_\psi &=&
\parbox{25mm} {
\begin{fmffile}{n19p}
   \begin{fmfgraph*}(70,50) \fmfkeep{theta1}
\fmfforce{(0.1w,0.5h)}{v1}
\fmfforce{(0.9w,0.5h)}{v2}
\fmf{dots_arrow, label=$j$}{v2,v1}
    \fmf{dashes_arrow, left=.7, label= $m$}{v1,v2}
    \fmf{dashes_arrow, left=.7, label= $n$}{v2,v1}
   \end{fmfgraph*}
\end{fmffile}
} 
[1+O(1/N^2)]
\nonumber\\ && 
=2 \prod_{l}\delta(\mu_l)\sum_{j,m,n}(\lambda_{j}+\lambda_{j}^2A_{j}^2
)
|V_{jn}^{m}|^2 |\Delta_{jn}^{m}|^2 \delta_{j+n}^{m}A_m^2A_n^2 \; [1+O(1/N^2)]
,
\label{b3}
\EEA
Summarising, we have
\BEA
J_2 = \prod_{l}\delta(\mu_l)\sum_{j,m,n}(\lambda_{j}+\lambda_{j}^2A_{j}^2
)
\left[|V_{mn}^{j}|^2 |\Delta_{mn}^{j}|^2 \delta_{m+n}^{j} +
2 |V_{jn}^{m}|^2 |\Delta_{jn}^{m}|^2 \delta_{j+n}^{m}
\right]
A_m^2A_n^2 \; [1+O(1/N)].
\EEA
Thus, we considered in detail the different terms involved
in $J_2$ and we found that the dominant contributions come
from the zero-valent graphs because they have more summation
indices involved. This turns out to be the general rule in both three-wave and
four-wave cases and it
allows one to simplify calculation by discarding a significant
number of graphs with non-zero valence.
Calculation of terms
$J_3$ to $J_5$ can be found in \cite{physd} and here we only present the
result,
\BE
J_3 = 4 \prod_l\delta(\mu_l)
\sum_{j,m,n}
\lambda_j
\left[ - |V_{mn}^j|^2\bar E(0,\omega_{mn}^j)\delta_{m+n}^jA_n^2+
|V_{jn}^m|^2 E(0,-\omega_{jn}^m)\delta_{j+n}^m (A_m^2 -A_n^2) 
\right] A_j^2 \times [1+O(1/N)],
\EE
$
J_4 = O(1)
$
(i.e. $J_4$ is order $N^2$ times smaller than  $J_2$ or  $J_3$) and 
\BE
J_5= 2 \prod_l\delta(\mu_l) \sum_{j\neq k,n}\lambda_j\lambda_k
\left[ 
-2 |V_{kn}^j|^2 \delta_{k+n}^j |\Delta_{kn}^j|^2 
+|V_{jk}^n|^2 \delta_{j+k}^n |\Delta_{jk}^n|^2 
\right]
A_j^2 A_n^2 A_k^2 \;\; [1+O(1/N)].
\EE

Using our results for $J_1 - J_5$ in (\ref{xt}) and (\ref{zx}) we have
\BEA
Z(T) - Z(0) &=& \e^2
 \sum_{j,m,n}(\lambda_{j}+\lambda_{j}^2 {\partial \over \partial \lambda_{j}})
\left[|V_{mn}^{j}|^2 |\Delta_{mn}^{j}|^2 \delta_{m+n}^{j} +
2 |V_{jn}^{m}|^2 |\Delta_{jn}^{m}|^2 \delta_{j+n}^{m}
\right]
{\partial^2 Z(0)\over \partial \lambda_{m} \partial \lambda_{n}} 
\nonumber \\
&& + 4\e^2 \sum_{j,m,n}
\lambda_j
\left[ - |V_{mn}^j|^2\bar E(0,\omega_{mn}^j)\delta_{m+n}^j
{\partial \over \partial \lambda_{n}}
+|V_{jn}^m|^2 E(0,-\omega_{jn}^m)\delta_{j+n}^m 
\left(
{\partial \over \partial \lambda_{m}}
- {\partial \over \partial \lambda_{n}} \right)
\right]  
{\partial Z(0) \over \partial \lambda_{j}}
\nonumber \\
&& + 2 \e^2 \sum_{j\neq k,n}\lambda_j\lambda_k
\left[ 
-2 |V_{kn}^j|^2 \delta_{k+n}^j |\Delta_{kn}^j|^2 
+|V_{jk}^n|^2 \delta_{j+k}^n |\Delta_{jk}^n|^2 
\right]
{\partial^3 Z(0)\over \partial \lambda_{j} \partial \lambda_{n}
\partial \lambda_{k}} \; +cc.
\label{discreteZ}
\EEA
Here partial derivatives with respect to $\lambda_l$ appeared because of 
the $A_l$ factors.
This expression is valid up to $O(\e^4)$
and $O(\epsilon^2/N)$ corrections.
Note that we still have not used any assumption about the statistics
of $A$'s. 
Let us now  $N \to \infty$ limit followed by $T \sim 1/\epsilon \to \infty$
(we re-iterate that this order of the limits is essential).
Taking into account that
$\lim\limits_{T\to\infty}E(0,x)= T (\pi
\delta(x)+iP(\frac{1}{x}))$, and
$\lim\limits_{T\to\infty}|\Delta(x)|^2=2\pi T\delta(x)$ and,
replacing $(Z(T) -Z(0))/T$ by $\dot Z$ we have
\BEA
\dot Z	&=&
4 \pi \e^2
 \int \big\{ (\lambda_{j}+\lambda_{j}^2 {\delta \over \delta \lambda_{j}})
\left[|V_{mn}^{j}|^2 \delta(\omega_{mn}^{j}) \delta_{m+n}^{j} +
2 |V_{jn}^{m}|^2 \delta(\omega_{jn}^{m}) \delta_{j+n}^{m}
\right]
{\delta^2 Z\over \delta \lambda_{m} \delta \lambda_{n}} 
\nonumber \\
&& + 2
\lambda_j
\left[ - |V_{mn}^j|^2 \delta(\omega_{mn}^j)\delta_{m+n}^j
{\delta \over \delta \lambda_{n}}
+|V_{jn}^m|^2 \delta(\omega_{jn}^m)\delta_{j+n}^m 
\left(
{\delta \over \delta \lambda_{m}}
- {\delta \over \delta \lambda_{n}} \right)
\right]  
{\delta Z \over \delta \lambda_{j}}
\nonumber \\
&& + 
2 \lambda_j\lambda_m
\left[ 
-2 |V_{mn}^j|^2 \delta_{m+n}^j \delta(\omega_{mn}^j)
+|V_{jm}^n|^2 \delta_{j+m}^n \delta(\omega_{jm}^n) 
\right]
{\delta^3 Z \over \delta \lambda_{j} \delta \lambda_{n}
\delta \lambda_{m}} \big\}\, dk_j dk_m dk_n.
\label{Zequat}
\EEA
Here variational derivatives appeared instead of partial derivatives because of 
the $N\to\infty$ limit.

Now we can observe that the
evolution equation for  $Z$ does not involve $\mu$ which
means that if initial  $Z$
 contains factor $
\prod_{l}\delta(\mu_l) $ it will be preserved at all time.
This, in turn, means that the phase factors $\{\psi
\}$ remain a set of statistically independent (of each each other and
of $A$'s) variables uniformly distributed on $S^1$. This is true with
accuracy $O(\e^2)$ (assuming that the $N$-limit is taken first, i.e.
$1/N \ll \e^2$) and this proves persistence of the first of the
``essential RPA'' properties. Similar result
for a special class of three-wave systems arising in the solid state physics
was previously obtained by Brout and Prigogine \cite{bp}.
This result is interesting 
because it has been obtained without any assumptions on the
statistics of the amplitudes $\{ A \}$ and, therefore, it is valid
beyond the RPA approach. It may appear useful in future for study of
fields with random phases but correlated amplitudes.


\subsection{Equation for the multi-mode PDF}


%
Taking the inverse Laplace transform of (\ref{Zequat}) we have the following
equation for the PDF,
\BE
\dot {\cal P} = - \int {\delta {\cal  F}_j \over \delta s_j} \, dk_j,
\label{peierls}
\EE
where ${\cal F}_j$ is a flux of probability in the space of the amplitude $s_j$,
\BEA
-{{\cal F}_j \over
4 \pi \e^2 s_j} 
&=&
\int
\big\{ 
(|V_{mn}^{j}|^2 \delta(\omega_{mn}^{j}) \delta_{m+n}^{j} +
2 |V_{jm}^{n}|^2 \delta(\omega_{jm}^{n}) \delta_{j+m}^{n}
)
s_n s_m {\delta {\cal P} \over \delta s_j}
\nonumber \\
&&
+2 {\cal P} (
|V_{jm}^{n}|^2 \delta(\omega_{jm}^{n}) \delta_{j+m}^{n}
- |V_{mn}^{j}|^2 \delta(\omega_{mn}^{j}) \delta_{m+n}^{j} 
)s_m
\nonumber \\
&&
+2
(|V_{jm}^{n}|^2 \delta(\omega_{jm}^{n}) \delta_{j+m}^{n}
-2|V_{mn}^{j}|^2 \delta(\omega_{mn}^{j}) \delta_{m+n}^{j} )
s_n s_m
{\delta {\cal P} \over \delta s_m}
 \big\} \, dk_m dk_n.
\label{flux}
\EEA
This equation is identical to the one originally obtained
by Peierls \cite{peierls} and later rediscovered
by Brout and Prigogine \cite{bp} in the context of
the physics of anharmonic crystals,
\BE
\dot  {\cal P} = {16 \pi \e^2 } \int |V_{mn}^{j}|^2 \delta(\omega_{mn}^{j}) \delta_{m+n}^{j}
\left[{\delta  \over \delta s} \right]_3 \left(s_j s_m s_n 
\left[{\delta \over \delta s} \right]_3 {\cal P} \right)  \, dk_j dk_m dk_n,
\label{peierls1}
\EE
where 
\BE
\left[{\delta  \over \delta s} \right]_3 = {\delta  \over \delta s_j} - {\delta
  \over \delta s_m} -{\delta  \over \delta s_n}.
\label{peierls2}
\EE
Zaslavski and Sagdeev  \cite{zs} were the first to study this
equation in the WT context. However, the  analysis of 
\cite{peierls,bp,zs}
was restricted 
to the interaction Hamiltonians of the ``potential energy'' type,
i.e. the ones that involve only the coordinates but not the momenta.
This restriction leaves aside a great many important WT systems,
e.g. the capillary, Rossby, internal and MHD waves.
Our result above indicates that the Peierls equation
 is also valid in the most general case of 3-wave systems.
Note that Peierls form (\ref{peierls1}) - (\ref{peierls2}) looks somewhat more
elegant and symmetric than (\ref{peierls}) - (\ref{flux}).
However, form  (\ref{peierls}) - (\ref{flux}) has advantage because it is 
in a continuity equation form. Particularly for steady state solutions, one
can immediately integrate it once and obtain,
\BE
{\cal F}\{s(k_l)\}  = \hbox{curl}_s A = \int \e_{lnm}  {\delta  A\{s(k_n)\}  \over \delta s(k_m)}
 \, dk_m dk_n,
\label{curlflux}
\EE
where $A$ is an arbitrary functional of $s(k)$ and
$\e_{lnm} $ is the antisymmetric tensor,
\BEA
\e_{lnm} = 1 & \hbox{for} & k_l > k_m > k_n \; \; \hbox{and cyclic permutations of}
\;\;\; l,m,n, \nonumber \\
\e_{lnm} = -1 & \hbox{for all other} & k_l \ne k_m \ne k_n, \nonumber \\
\e_{lnm} = 0 & \hbox{if } & k_l=k_m , k_l= k_n \; \hbox{or} \; k_n=k_m. \nonumber
\EEA
In the other words, probability flux $F$ can be an arbitrary solenoidal field
in the functional space of  $s(k)$.  One can see that (\ref{curlflux}) is a
first order equation with respect to the $s$-derivative. 
Special cases of the steady solutions are  the zero-flux and the constant-flux
solutions which, as we will see later correspond to a Gaussian and
intermittent wave turbulence respectively.

Here we should again emphasise importance of the taken order
of limits, $N \to \infty$ first and $\e \to 0$ second.
Physically this means that the frequency resonance is broad
enough to cover great many modes. Some authors, e.g. \cite{peierls,bp,zs},
leave the sum notation in the PDF equation even after the
$\e \to 0$ limit taken giving $\delta(\omega_{jm}^{n})$.
One has to be careful interpreting such formulae because
formally the RHS is nill in most of the cases because 
there may be no exact resonances between the discrete $k$ modes
(as it is the case, e.g. for the capillary waves). In real finite-size
physical systems, this condition means that the wave amplitudes, although
small, should not be too small so that the frequency broadening is sufficient
to allow the resonant interactions. Our 
functional integral notation is meant to indicate that 
the $N \to \infty$ limit has already been taken.

\section{Evolution of statistics of four-wave systems}

%

In this section we are going to calculate the four-wave analog of
PBP equation. These calculations are similar in spirit to
those presented in the orevious section, but with the four-wave equation
of motion (\ref{FourWaveEquationOfMotionA}). The calculation will be slightly
more
involved due to the frequency renormalisation, but it will be significantly
simplified by our knowledge that 
(for the same reason as for three-wave case) all
$\mu$ dependent terms will not contribute to the final result. Indeed,
any term with nonzero $\mu$ would have less summations and therefore
will be of lower order.  On the diagrammatic language that would mean
that all diagrams with non-zero valence can be discarded, as it was explained
in the previous section. Below, we will only keep the leading order terms
which are $\sim N^2$ and wich correspond to the zero-valence diagrams.
Thus, to calculate $Z$ we start with the  $J_1-J_5$ terms
 (\ref{j1}),(\ref{j2}),(\ref{j3}),(\ref{j4}),(\ref{j5}) in which we put $\mu=0$ and 
substutute into them the values of $a\one$ and
$a\two$ from (\ref{FirstIterate1}) and (\ref{SecondIterate1}).
 For the terms proportional to $\epsilon$ we have
\begin{eqnarray}
\langle a_j\one\bar a_j^{(0)}\rangle_\psi= \nonumber \\
- i \sum_{\alpha\mu\nu} W^{j\alpha}_{\mu\nu}
\langle \bar a_j \bar
a_\alpha a_\mu a_\nu\rangle_\psi
\delta^{j\alpha}_{\mu\nu} \Delta^{j
\alpha}_{\mu\nu} + i \Omega_j \langle |a_j|^2 \rangle_\psi  T\nonumber\\
=- 2 i \sum_{\alpha} W^{j\alpha}_{j\alpha}
A_j^2 A_\alpha^2 \cdot T
+ i \Omega_j A_j^2 T.
\label{EpsilonTerms}
\end{eqnarray}
where we have used the fact that $\Delta(0)=T$.
We see that our choice of the frequency renormalisation
(\ref{FrequencyRenormalization}) 
makes the contribution of $\langle a_j\one\bar a_k^{(0)}\rangle_\psi$
equal to zero, and this is the main reason for making this choice.
Also, this choice of the frequency correction simplifies the $\e^2$ order and,
most importantly, eliminates from them the $T^2$ cumulative growth, e.g.
\BE
\langle
(a_j\one\bar a_j^{(0)})^2
\rangle_\psi=
-2 W^{jj }_{jj } A_j^6 T^2 \Omega_j
-A_j^4 \Omega_j^2 T^2 - 4 \sum_\alpha(W^{j \alpha}_{j \alpha}
A_j A_\alpha T )^2\to 0 {\rm \ in \; the \ }N\to\infty {\rm \ limit. \ }\nonumber
\EE
Finally we get,
\BEA
J_1 &=& 0,\CR
J_2 &=& \sum_{j\alpha\mu\nu}\lambda_j(1 + \lambda_j
A_j^2)|W_{\mu\nu}^{j\alpha}|^2|\Delta_{\mu\nu}^{j\alpha}|^2
\delta_{\mu\nu}^{j\alpha} A_\mu^2 A_\alpha^2 A_\nu^2,
  \CR
J_3 &=& 2\sum_{j\alpha\mu\nu} \lambda_j|W_{\mu\nu}^{j\alpha}|^2
\delta_{\mu\nu}^{j\alpha} A_j^2 A_\nu^2 (A_\mu^2
-2A_\alpha^2 ) E(0,\tilde {\omega}_{\mu\nu}^{j\alpha}),\CR
J_4 &=&  0,\CR
J_5 &=& \sum_{j\neq
 k}\lambda_j\lambda_k\left[-2|W_{k\nu}^{j\alpha}|^2
 |\Delta_{k\nu}^{j\alpha}|^2\delta_{k\nu}^{j\alpha} A_\alpha^2 +
 |W_{\mu\nu}^{jk}|^2 |\Delta_{\mu\nu}^{jk}|^2\delta_{\mu\nu}^{jk}A_\mu^2
 \right]A_j^2 A_k^2 A_\nu^2.
\EEA
By putting everything together in (\ref{zx}) and (\ref{xt}) and explointing the symmeteries
introduced by the summation variables, we finally obtain  in
$N \to \infty$ limit followed by $T   \to \infty$
 \BE
\dot Z =
4 \pi \e^2 \int |W^{jl}_{nm}|^2\delta(\tilde\omega^{jl}_{nm})
\delta^{jl}_{nm}
  \lambda_j 
\frac{\partial}{\partial \lambda_j} \frac{\partial}{\partial \lambda_l}
\frac{\partial}{\partial \lambda_n} \frac{\partial}{\partial \lambda_m}
\left[ (\lambda_j + \lambda_l - \lambda_m - \lambda_n)
 Z \right] \, dk_j  dk_l dk_m dk_n.
\label{dotZ}
\EE
By applying to this equation the inverse Laplace transform, we
get a four-wave analog of the PBP equation for the $N$-mode PDF:
\BE
\dot  {\cal P} = { \pi \e^2 } \int |W^{jl}_{nm}|^2\delta(\tilde\omega^{jl}_{nm})
\delta^{jl}_{nm}
\left[{\delta  \over \delta s} \right]_4 \left(s_j s_ls_m s_n 
\left[{\delta \over \delta s} \right]_4 {\cal P} \right)  \, dk_j  dk_l dk_m dk_n,
\label{peierls4}
\EE
where 
\BE
\left[{\delta  \over \delta s} \right]_4 = {\delta  \over \delta s_j}+{\delta  \over \delta s_l} - {\delta
  \over \delta s_m} -{\delta  \over \delta s_n}.
\label{brac4}
\EE
This equation can be easily written in the continuity equation form
(\ref{peierls}) with the flux given in this case by
\BE
{\cal F}_j = -{4 s_j \pi \e^2 } \int |W^{jl}_{nm}|^2\delta(\tilde\omega^{jl}_{nm})
\delta^{jl}_{nm}
s_ls_m s_n 
\left[{\delta \over \delta s} \right]_4 {\cal P}   \,  dk_l dk_m dk_n,
\label{peierls4}
\EE

\section{Justification of RPA}

Variables $s_j$ do not separate 
in the above equation for the PDF.
Indeed, substituting 
\BE
{\cal P}^{(N,a)} =
 P^{(a)}_{j_1}  P^{(a)}_{j_2} \dots  P^{(a)}_{j_N} \; 
\label{pure}
\EE
into the discrete version of (\ref{flux}) we
see that it turns into zero on the thermodynamic solution
with $P^{(a)}_{j} = \omega_j \exp(-\omega_j s_j)$.
However, it is not
zero for the one-mode PDF  $P^{(a)}_{j}$
corresponding to the cascade-type Kolmogorov-Zakharov (KZ)
spectrum $n_j^{kz}$, i.e.
 $P^{(a)}_{j} = (1/n_j^{kz}) \exp(-s_j /n_j^{kz})$ (see the Appendix), 
nor it is likely to be zero for any other
PDF of form (\ref{pure}).
This means that, even initially independent, 
the amplitudes will correlate with each other
at the nonlinear time. Does this mean that the existing 
WT theory, and in particular the kinetic equation, is 
invalid?

To answer to this question let us differentiate the discrete version
of the equation (\ref{Zequat}) with respect to $\lambda$'s to get
equations for the amplitude moments. We can easily see that
\BE
\partial_t \left(\langle A_{j_1}^2 A_{j_2}^2 \rangle 
-\langle A_{j_1}^2 \rangle  \langle A_{j_2}^2 \rangle \right) =
O(\e^4) \quad (j_1, j_2 \in {\cal B}_N) 
\label{split}
\EE  
if $\langle A_{j_1}^2 A_{j_2}^2 A_{j_3}^2 \rangle =
\langle A_{j_1}^2 \rangle  \langle A_{j_2}^2 \rangle  \langle
A_{j_3}^2 \rangle $ (with the same accuracy)
at $t=0$.
Similarly, in terms of PDF's 
\BE
\partial_t \left(P^{(2,a)}_{j_1, j_2} (s_{j_1}, s_{j_2}) 
- P^{(a)}_{j_1}(s_{j_1}) P^{(a)}_{j_2}(s_{j_2})  \right) =
 O(\e^4) \quad (j_1, j_2  \in {\cal B}_N)
\EE  
if $P^{(4,a)}_{j_1, j_2, j_3, j_4} (s_{j_1}, s_{j_2}, s_{j_3}, s_{j_4}) =
P^{(a)}_{j_1}(s_{j_1}) P^{(a)}_{j_2}(s_{j_2}) P^{(a)}_{j_3}(s_{j_3})  P^{(a)}_{j_4}(s_{j_4}) 
$ at $t=0$.
Here $P^{(4,a)}_{j_1, j_2, j_3, j_4} (s_{j_1}, s_{j_2}, s_{j_3}, s_{j_4}) $,
$P^{(2,a)}_{j_1, j_2} (s_{j_1}, s_{j_2}) $ and 
$P^{(a)}_{j} (s_{j}) $ 
are
the four-mode,
two-mode and one-mode PDF's obtained
from $\cal P$ by integrating out all but 3,2 or  1
arguments respectively.
One can see that, with a $\e^2$ accuracy,
the Fourier modes will remain independent of each other
in any pair over the nonlinear time if they were
independent in every triplet at $t=0$.

Similarly, one can show that the modes will remain
independent over the nonlinear time
in any subset of $M<N$ modes with accuracy $M/N$
(and $\e^2$) if they were initially independent in
every subset of size $M+1$. Namely
\BEA
P^{(M,a)}_{j_1, j_2, \dots , j_M} (s_{j_1}, s_{j_2}, s_{j_M}) 
- P^{(a)}_{j_1}(s_{j_1}) P^{(a)}_{j_2}(s_{j_2} ) \dots
P^{(a)}_{j_M}(s_{j_M} )   =
O(M/N) + O(\e^2) \nonumber \\
\quad (j_1, j_2, \dots, j_M \in {\cal B}_N)
\EEA  
if $P^{(M+1,a)}_{j_1, j_2, \dots, j_{M+1}} =
P^{(a)}_{j_1}  P^{(a)}_{j_2} \dots P^{(a)}_{j_{M+1}}
$ at $t=0$.

Mismatch $O(M/N)$ arises from some terms in the ZS equation with
coinciding indices $j$. For $M=2$ there is only one such term
in the $N$-sum and, therefore, the corresponding error is
$O(1/N)$ which is much less than $O(\epsilon^2)$ (due to the
order of the limits in $N$ and $\epsilon$).
However, the number of such terms grows as $M$ and the error
accumulates to  $O( M/N)$ which can greatly exceed $O(\epsilon^2)$
for sufficiently large $M$.

We see that the accuracy with which the modes remain
independent in a subset is worse for larger
subsets and that the independence property is
completely lost for subsets  approaching in size
the entire set, $M \sim N$.
One should not worry too much about this loss
because $N$ is the biggest parameter in the 
problem (size of the box) and the modes
will be independent in all
$M$-subsets no matter how large.
Thus, the statistical objects
involving any {\em finite} number of modes
are factorisable as products of the one-mode
objects and, therefore, the WT theory  reduces to 
considering the one-mode objects. 
This results explains why we re-defined RPA in its
relaxed ``essential RPA'' form.
Indeed, in this form RPA is sufficient for the WT
closure and, on the other hand, it remains valid over the nonlinear
time. In particular, only property (\ref{split}) is needed,
as far as the amplitude statistics is concerned, for deriving
the 3-wave kinetic equation, and this fact validates this equation
and all of its solutions, including the KZ spectrum which plays an
important role in WT.

The situation were modes can be considered as independent when
taken in relatively small sets but should be treated as
 dependent in the context
of much  larger sets is not so unusual in physics. Consider for
example a distribution of electrons and ions in plasma.
The full $N$-particle distribution function in this case satisfies
the Liouville  equation which is, in general, not a separable
equation. In other words, the $N$-particle distribution function
cannot be written as a product of $N$ 
one-particle distribution functions. However, an $M$-particle
distribution can indeed be represented as 
a product of $M$ 
one-particle distributions if $M \ll N_D$ where $N_D$ is the number
of particles in the Debye sphere. We see  an interesting transition
from a an individual to collective behaviour when the number
of particles  approaches $N_D$. In the special case of the one-particle
function we have here the famous mean-field Vlasov equation which is
valid up to  $O(1/N_D)$ corrections (representing particle collisions).

\section{One-mode statistics}

We have established above that 
the one-point statistics is at the heart of the WT theory.
All one-point statistical objects can be derived from the one-point
amplitude generating function, 
$$Z_a (\lambda_j) = \left< e^{\lambda_j A_j^2} \right> $$ 
which can be obtained from the $N$-point $Z$ by taking all $\mu$'s
and all $\lambda$'s, except for $\lambda_j$, equal to zero.
Substituting such values to (\ref{Zequat}) (for the three-wave case)
or to (\ref{dotZ}) (four-wave) we get the following
equation for $Z_a$,
\begin{equation}
\frac{\partial Z_a}{\partial t} = \lambda_j \eta_j Z_a +(\lambda_j^2
\eta_j - \lambda_j \gamma_j) \frac{\partial Z_a}{\partial \lambda_j},
\label{za}
\end{equation}
where for the three-wave case we have:
\BEA \eta_j  = 4 \pi \epsilon^2 \int 
\left(|V^j_{lm}|^2 \delta^j_{lm}  \delta(\omega^j_{lm})
+2 |V^m_{jl}|^2 \delta^m_{jl}  \delta(\omega^m_{jl} )
\right)  n_{l} n_{m}
\, d { k_l} d { k_m} ,  \label{RHO} \\
 \gamma_j =
8 \pi \epsilon^2 \int  
\left(
|V^j_{lm}|^2 \delta^j_{lm} \delta(\omega^j_{lm}) n_{m}
 +|V^m_{jl} |^2 \delta^m_{jl} \delta(\omega^m_{jl}) (n_{l}- n_{m})
\right) \, d { k_l} d { k_m}  .  \label{GAMMA}
\EEA
and in the four-wave case:
\begin{eqnarray}
\eta_j &=& 4 \pi \epsilon^2 \int
|W^{jl}_{nm}|^2 \delta^{jl}_{nm}
\delta(\omega^{jl}_{nm}) n_l n_m n_n \,  d { k_l} d { k_m}  d { k_n,} 
\label{RHO1} \\
\gamma_j &=&
4 \pi \epsilon^2 \int  |W^{jl}_{nm}|^2 \delta^{jl}_{nm}
\delta(\omega^{jl}_{nm})
 \Big[ n_l (n_m + n_n) - n_m n_n\Big] \,  d { k_l} d { k_m}  d { k_n.}
 \label{GAMMA1}
\end{eqnarray}
Here we introduced the wave-action spectrum,
\BE
n_j = \langle A_j^2 \rangle.
\EE 
Differentiating (\ref{za}) with respect to $\lambda$  we get an equation
for the  moments
$ M^{(p)}_j = \langle A_j^{2p} \rangle $:
\BE
\dot M^{(p)}_j = -p \gamma_j M^{(p)}_j +
p^2 \eta_j M^{(p-1)}_j.\label{MainResultOne} 
\EE
which, for $p=1$ gives the standard kinetic equation,
\BE
 \dot n_j = - \gamma_j n_j + \eta_j . 
\label{ke}
\EE

First-order PDE (\ref{za}) can be easily solved by the method of characteristics.
Its steady state solution  is 
\BE
 Z={1 \over 1 -\lambda n_k} 
\EE
which corresponds to the Gaussian values of momenta 
\BE
M^{(p)} = p! n_k^p.
\EE
However, these solutions are invalid at small $\lambda$ and high $p$'s
because large amplitudes $s=|a|^2$, for which nonlinearity is not weak,
 strongly contribute in these cases. Because of the integral nature
of definitions of $M^{(p)}$ and $Z$ with respect to the $s=|a|^2$,
the ranges of amplitudes where WT is applicable are mixed with,
and contaminated
by, the regions where WT fails. Thus, to clearly separate these regions
it is better to work with quantities which are local in $s=|a|^2$,
in particular the one-mode probability distribution $P_a$.
Equation for the one-mode PDF can be obtained by applying the inverse Laplace
transform to  (\ref{za}). This gives:
\begin{equation}
{\partial P_a \over \partial t}+  {\partial F \over \partial s_j}  =0,
 \label{pa}
\end{equation}
 with $F$ is a probability flux in the s-space,
\begin{equation}
F=-s_j (\gamma P_a +\eta_j {\delta P_a \over \delta s_j}).
\label{flux1}
\end{equation}
Let us consider the steady state solutions, $\dot P_{a} =0$ so that
\BE
-s(\gamma P+\eta\partial_{s}P) = F= \hbox{const}. \label{ff}
\EE
Note that in the steady state $\gamma /\eta = n$ which follows from
kinetic equation (\ref{ke}).
The general solution to (\ref{ff})  is 
\BE
P=P_{hom} + P_{part}
\EE
where 
\BE
P_{hom} = \hbox{const} \, \exp{(-s/n)}
\EE
 is the general solution to the  homogeneous
equation (corresponding to $F=0$) 
and $P_{part}$ is a particular solution,
\BE
P_{part} = -({F}/{\eta}) Ei({s}/{n}) \exp{(-s/n)}, 
\EE
where $Ei(x)$ is the integral exponential function.

At the tail of the PDF, $s \gg n_k$, the solution can be represented as series
in $1/s$,
\begin{equation}
P_{part} = - \frac{F}{s\gamma}-\frac{\eta F}{(\gamma s)^2}+
\cdots\label{Ppart}.
\end{equation}
Thus, the leading order asymptotics of the finite-flux solution is
$1/s$ which 
decays much slower than the exponential (Rayleigh) part $P_{hom}$ and, therefore, 
describes strong intermittency.

Note that if the weakly nonlinearity 
assumption was valid uniformly to $s=\infty$ then we had to
put $F=0$ to ensure positivity of $P$ and the convergence of its normalisation,
$\int P \, ds =1$. In this case $P=P_{hom} = n \, \exp{(-s/n)}$ which
is a pure Rayleigh distribution corresponding to the Gaussian wave field.
However, WT approach fails for  for the
amplitudes $s \ge s_{nl}$ for which the nonlinear time is of the same order
or less than the linear wave period and, therefore, we can expect
a cut-off of $P(s)$ at $s = s_{nl}$. Estimate for the value of
$s_{nl}$ can be obtained from the dynamical
equation (\ref{FourWaveEquationOfMotionB}) by balancing the linear and
nonlinear
terms and assuming that 
if the wave amplitude at some $k$ happened to be of the critical
value $s_{nl}$ then it will also be of similar value for a range of $k$'s
of width $k$ (i.e. the $k$-modes are strongly correlated when the amplitude is
close to critical). 
This gives:
\BE
 s_{nl}= \omega/\epsilon W k^2 
\EE
This cutoff can be viewed as a wavebreaking process which does not allow
wave amplitudes to exceed their critical value, $P(s) =0$ for $s > s_{nl}$.
Now the normalisation condition can be satisfied for the finite-flux
solutions. Note that in order for our analysis to give the correct description
of the PDF tail, the nonlinearity must remain weak for the tail, which means
that
the breakdown happens far from the PDF core $s \ll s_{nl}$. When $s_{nl} \sim
n$ one has a strong breakdown predicted in \cite{biven} which is hard to
describe rigorously due to strong nonlinearity.

\begin{figure}[b]
  \begin{center}
  \includegraphics[width=.5\textwidth]{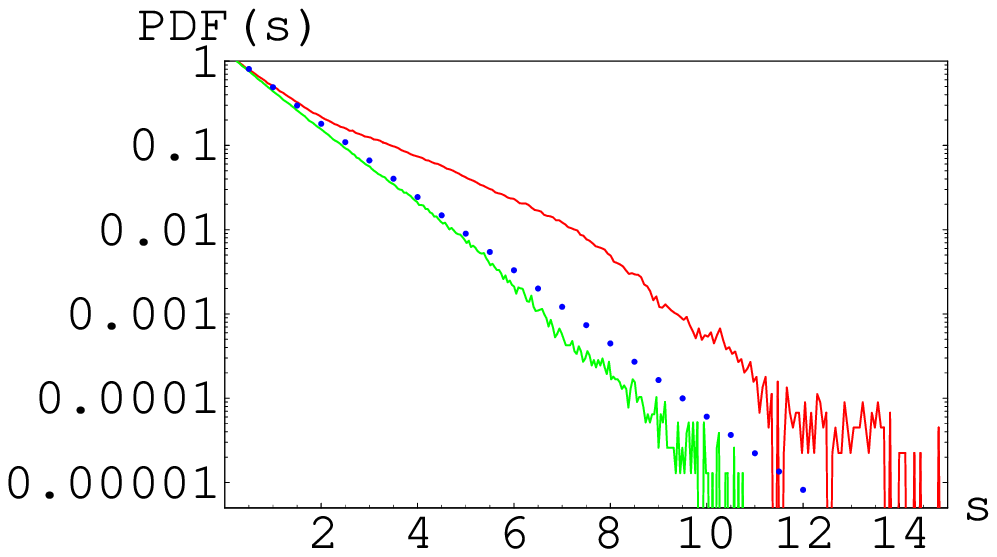}
  \end{center}
  \caption[]{One-mode amplitude PDF's at two wavenumbers: upper curve at $k_1$ 
and the lower curve at and  $k_2$
such that  $k_1>k_2$. Dashed line corresponds to the Rayleigh distribution.}
  \label{fig1}
\end{figure}

Depending on the position in the wavenumber space, the flux $F$ can be
either positive or negative. As we discussed above, $F<0$ results in an 
enhanced probability of large wave amplitudes with respect to the Gaussian
fields. Positive $F$ mean depleted probability and correspond to the
wavebreaking value  $s_{nl}$ which is closer to the PDF core.
When $s_{nl}$ gets into the core, $s_{nl} \sim n$ one reaches the wavenumbers
at which the breakdown is strong, i.e. of the kind considered in \cite{biven}.
Consider for example the water surface gravity waves. Analysis of \cite{biven}
predicts strong breakdown in the high-$k$ part of the energy cascade range.
According to our picture, these high wavenumbers correspond to the highest
positive values of $F$ and, therefore, the most depleted PDF tails with
respect to the Gaussian distribution. When one moves away from this region
toward lower $k$'s, the value of $F$ gets smaller and, eventually, changes the
sign leading to enhanced PDF tails at low $k$'s.
This picture is confirmed by the direct numerical simulations of the water
surface equations reported in
\cite{cln} the results of which are shown in figure \ref{fig1}.

 This figure shows
that at a high $k$ the PDF tail is depleted with respect to the Rayleigh
distribution, whereas at a lower $k$ it is enhanced which corresponds to
intermittency at this scale. 
Similar conclusion that the gravity wave turbulence is intermittent at low
rather than high wavenumbers was reached on the basis of numerical simulations
in \cite{yoko}.
To understand the flux reversal leading to intermittency appears in the
one-mode statistics, one has to
consider fluxes in the multi-mode phase space which will be done in the next section.

\section{Intermittency and the multi-mode probability vortex.}

\begin{figure}[b]
  \begin{center}
  \includegraphics[width=.4\textwidth]{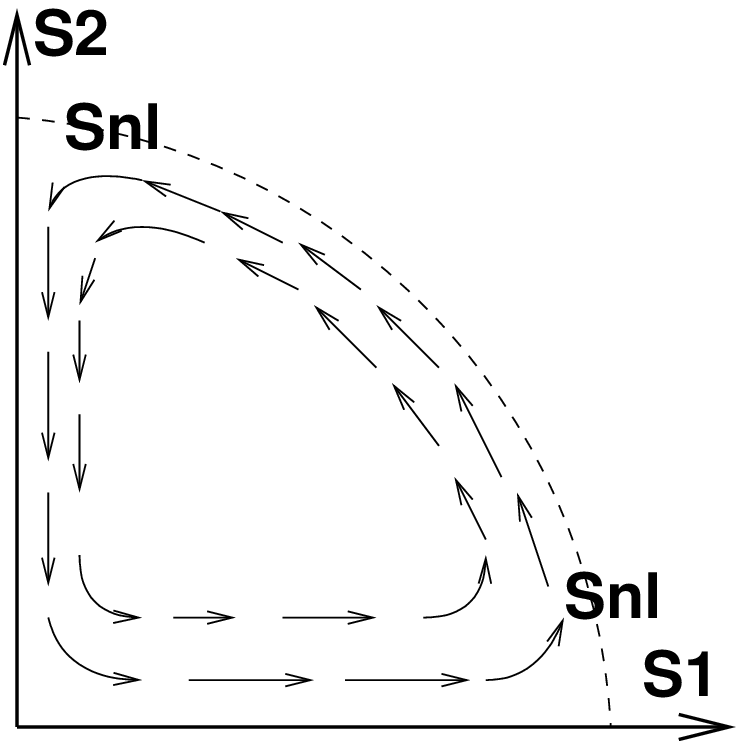}
  \end{center}
  \caption[]{Projection of the probability flux vortex on a $(s_1,s_2)$ plane
  where $s_1$ and  $s_2$ are the amplitudes at wavenumbers $k_1$ and  $k_2$
such that  $k_1>k_2$.}
  \label{fig2}
\end{figure}

In the previous section, we established that one-mode PDF's can deviate
from the Rayleigh distributions if the flux of probability in the amplitude
space is not equal to zero. However, in the full $N$-mode amplitude space,
the flux lines cannot originate or terminate, i.e. there the probability 
``sources'' and ``sinks'' are impossible,  see (\ref{curlflux}).
Even adding forcing or dissipation into the dynamical equations does not
change this fact because this can only modify the expression for the flux (see
the Appendix) but it 
cannot change the PDF continuity equation  (\ref{peierls}).
Thus, presence of the finite flux for the
one-point PDF's corresponds to deviation of the flux lines from the straight
lines in the   $N$-mode amplitude space. The global structure of such
a solution in the  $N$-mode space corresponds to a $N$ dimensional
probability vortex. This probability vortex is illustrated in figure ? which
sketches its projection onto a a 2D plane corresponding to one low-wavenumber and one
high-wavenumber amplitudes. Taking 1D sections of this vortex one observes
a positive one-mode flux at high $k$ and a negative 
 one-mode flux at low $k$, in accordance with the numerical observations of
 figure \ref{fig2}.
 We should say, however, that existence of the probability vortex
 solutions, although consistent with numerics, remains hypothetical and
 further work needs to be done to find solutions of  (\ref{curlflux}) with
 non-zero curl.

\section{Discussion.}

In this paper, we reviewed recent work in the field of Wave Turbulence devoted
to study non-Gaussian aspects of the wave statistics, intermittency,
validation of the phase and amplitude randomness, higher
spectral moments and fluctuations. We also presented some new results,
particularly derivation of 
the analog of the Peierls-Brout-Prigogine equation for the four-wave systems.
The wavefileds we dealt with are,
generally, characterised by non-decaying correlations along certain directions
in the coordinate space. These fields are typical for WT because, due to weak
nonlinearity, wavepackets preserve identity over long distances. One of the
most common examples of such long-correlated fields is given by the typical
initial condition in numerical simulations where the phases are random but the
amplitudes are chosen to be deterministic. We showed that wavefields can
develop enhanced probabilities of high amplitudes at some wavenumbers which
corresponds to intermittency. Simultaneously, at other wavenumbers, the
probability of high amplitudes can be depleted with respect to Gaussian
statistics. We showed that both PDF tail enhancement and its depletion related
to presence of a probability flux in the amplitude space (which is positive for
depletion and negative for the enhancement). We speculated that the $N$-dimensional
space of $N$ amplitudes, these fluxes correspond to an $N$-dimensional
probability vortex. We argued that presence of such vortex is prompted by
non-existence of a zero-amplitude-flux solution corresponding to the KZ spectrum with
de-correlated amplitudes. Finding such a probability vortex solution
analytically remains a task for future. 

\section{Appendix: Wave Turbulence with sources and sinks}

One of the central discoveries in Wave Turbulence was the power-law
Kolmogorov-Zakharov (KZ)
spectrum, $n^{kz} \sim k^\nu$, which realise themselves in presence of the energy sources and sinks
separated by a large inertial range of scales. The exponent $\nu$ depends on
the
scaling properties of the interaction coefficient and the frequency. Most of the previous WT
literature
is devoted to study of KZ spectra and a good review of these works can be
found in \cite{ZLF}. We are not going review these studies here, but instead
we
are going to  find out how adding such energy sources and sinks will this modify
the evolution equations for the statistics.
Instead of  the Hamiltonian equation (\ref{HamiltonianEquationOfMotion}) let
us consider
\BEA
i\dot c_l &=&\frac{\partial {\cal H}}{\partial \bar c_l} + i \tilde \gamma_l c_l, 
\label{HamiltonianEquationOfMotion1}
\EEA
where $\tilde \gamma_l$ describes sources and sinks of the energy, e.g. due to
instability and viscosity respectively. Easy to see that this linear term will
not change the structure of the $N$-mode PDF equation (\ref{peierls}) but it
will lead to re-definition of the flux:
\BE
{\cal F}_j \to {\cal F}_j - \tilde \gamma_j s_j P.
\EE
In the one-mode equations, this simply means renormalisation 
\BE
\gamma \to \gamma - \tilde \gamma.
\EE
Thus we arrive at a simple message that the energy sources and sinks do not
produce any ``sources'' or ``sinks'' for the flux of probability.

In the inertial range, there is no flux modification and one can easily find
 $F=0$ solution of (\ref{pa}) for the one-mode PDF,
\BE
 P^{(a)}_{j} = (1/n_j^{kz}) \exp(-s_j /n_j^{kz}),
\EE 
where $n_j^{kz}$ is the KZ spectrum (solving the kinetic equation in the
 inertial range). However, it is easy to check by substitution that
the product of such one-mode PDF's, ${\cal P} = \Pi_j (1/n_j^{kz}) \exp(-s_j
 /n_j^{kz})$, is not an exact solution to the multi-mode equation
(\ref{peierls}). 
Thus, there have to be corrections to this expression related either to a
 finite flux, an amplitude correlation or both.


\end{document}